\documentclass[usenatbib]{mnras}
\usepackage{graphics}
\usepackage{rotating}
\usepackage{amsmath}
\usepackage{breqn}
\usepackage{amssymb}
\PassOptionsToPackage{pdfpagelabels=false}{hyperref}
\usepackage{multirow}
\usepackage{booktabs}

\title{On the small scale clustering of quasars: constraints from the MassiveBlack II simulation}
\author[Bhowmick et al.]{
Aklant K. Bhowmick$^{1}$,
Tiziana DiMatteo$^{1}$,
Sarah Eftekharzadeh$^{2,3}$,
Adam D. Myers$^{4}$
\\
$^{1}$McWilliams Center for Cosmology, Dept. of Physics, Carnegie Mellon University,
Pittsburgh PA 15213, USA\\
$^{2}$Department of Physics, Southern Methodist University, 3215 Daniel Ave., Dallas, TX 75225\\
$^{3}$Department of Physics \& Astronomy, University of Utah, Salt Lake City, UT 84112\\
$^{4}$Department of Physics \& Astronomy, University of Wyoming, 1000 University Ave., Laramie, WY, 82071\\
}
\begin{document}
\maketitle

\begin{abstract}
We examine recent high-precision measurements of small-scale quasar clustering (at $z\sim0.5-2$ on scales of $\sim25~\mathrm{kpc/h}$) from the SDSS in the context of the MassiveBlackII (MBII) cosmological hydrodynamic simulation and conditional luminosity function (CLF) modeling. At these high luminosities ($g < 20.85$ quasars), the MBII simulation volume ($100~\mathrm{cMpc}/h$ comoving boxsize) has only 3 quasar pairs at distances of $1-4$ Mpc.
The black-hole masses for the pairs range between $M_{bh}\sim1-3\times 10^{9}~M_{\odot}/h$ and the quasar hosts are haloes of $M_h\sim1-3\times10^{14}~M_{\odot}/h$. Such pairs show signs of recent major mergers in the MBII simulation. By modeling the central and satellite AGN CLFs as log-normal and Schechter distributions respectively (as seen in MBII AGNs), we arrive at CLF models which fit the simulation predictions and observed luminosity function and the small-scale clustering measured for the SDSS sample. The small-scale clustering of our mock quasars is well-explained by central-satellite quasar pairs that reside in $M_h>10^{14}~M_{\odot}/h$ dark matter haloes. For these pairs, satellite quasar luminosity is similar to that of central quasars. 
Our CLF models imply a relatively steep increase 
in the maximum satellite luminosity, $L^*_{\mathrm{sat}}$,
in haloes of $M_h>10^{14}~M_{\odot}/h$ with associated larger values of $L^*_{\mathrm{sat}}$ at higher redshift. This leads to increase in the satellite fraction that manifests itself in an enhanced clustering signal at $\lesssim$ 1 Mpc/h. For the ongoing eBOSS-CORE sample, we predict $\sim 200-500$ quasar pairs at $z\sim1.5$ (with $M_h \gtrsim10^{13}~M_{\odot}/h$ and $M_{bh} \gtrsim10^{8}~M_{\odot}/h$) at $\sim25~\mathrm{kpc}$ scales. Such a sample would be $\gtrsim10$ times larger than current pair samples.
\end{abstract}


\section{Introduction}
Small scale clustering measurements for AGNs/quasars have been of significant interest over the last two decades as they may constrain signatures of the physical processes that trigger AGN activity, such as galaxy mergers \citep{2005Natur.433..604D,1999MNRAS.309..836M,1999ApJ...510..590K}.
Several works \citep{2000AJ....120.2183S,2006AJ....131....1H,2007ApJ...658...99M, 2008ApJ...678..635M, 2012MNRAS.424.1363K,2016AJ....151...61M,2017MNRAS.468...77E} over the last 20 years have measured the small-scale clustering of quasars, mainly from the SDSS and 2dF-QSO surveys, at scales ranging from $ \sim 10~\mathrm{kpc}$ to $\sim1$ Mpc. These works typically measure strong clustering at small scales~($\lesssim 200~\mathrm{kpc}$), which could be attributed to galaxy mergers as possible triggers of quasar activity. 

Cosmological hydrodynamic simulations are valuable tools to study AGN clustering. Theoretical predictions for AGN clustering and luminosity evolution have been made using \texttt{ILLUSTRIS}~\citep{2017MNRAS.466.3331D} and \texttt{MassiveBlackII}~\citep[][hereafter MBII]{2015MNRAS.450.1349K} and they report broad agreement with observational measurements \citep{2005MNRAS.356..415C,2006MNRAS.371.1824P,2006ApJ...638..622M,2009ApJ...697.1656S,2012MNRAS.424..933W,2015MNRAS.453.2779E}. However, simulated AGNs are significantly fainter (with bolometric luminosities $10^{41}\lesssim L\lesssim 10^{44}~\mathrm{ergs/sec}$) than the observed quasar samples~($L\gtrsim 10^{45}~\mathrm{ergs/sec}$). These bright quasars are too rare for their clustering properties to be directly probed by hydrodynamic simulations \citep{2015MNRAS.450.1349K,2015MNRAS.446..521S,2015A&C....13...12N,2017MNRAS.467.4739K}. In general, this is because the volumes of such simulations~(boxsizes are $100~\mathrm{cMpc}/h$ for MBII and \texttt{Horizon AGN}, and $\sim 75~\mathrm{cMpc}/h$ for \texttt{ILLUSTRIS} and \texttt{EAGLE}) are limited due to the computational demand of implementing `full physics'. However, analytical models such as Halo Occupation Distribution~(HOD) or Conditional Luminosity Function~(CLF) modeling can be used to supplement simulations. This mitigates volume limitations and allows simulations to probe the implications of clustering measurements at small scales. 

HOD and CLF modeling offer predictions for the properties of dark matter haloes based on the observed clustering of the targets that haloes host. This has been applied extensively to galaxies \citep[and references therein]{2007ApJ...667..760Z,2016MNRAS.459.3040G,2018PASJ...70S..11H} as well as AGNs \citep[and references therein]{2013ApJ...779..147C,2017MNRAS.464..613B,2018MNRAS.477...45M,2018arXiv181205760E}. In this work, we seek a comparison between theory and observations for the quasar population with $g<20.85$ and $0.43\lesssim z \lesssim2.2$. Here we build a quasar CLF by extrapolating from the properties of AGN in the simulations.
These models allow us to effectively populate AGNs in more massive dark-matter haloes and, subsequently, to make clustering predictions for objects that are too rare for simulations to directly probe. 

The recent work by \citet[hereafter E17]{2017MNRAS.468...77E} makes the most precise measurement of quasar clustering at $\sim25~\mathrm{kpc}/h$ scales to date over redshifts of $0.4\lesssim z\lesssim 2.3$. In this work, we use our cosmological hydrodynamic simulation, MassiveBlack II \citep{2015MNRAS.450.1349K}, to probe the small-scale clustering of AGNs and compare our results to their measurements. To bridge the gap between faint ($g\gtrsim 23$) simulated AGNs and observed bright quasars ($g\lesssim 20.85$), we construct CLF models to predict the one-halo clustering of these quasars. 

We then focus on the clustering signal at scales probed by observations ($\sim25~\mathrm{kpc}$ proper separations) and finally arrive at a CLF model which predicts a small scale quasar  clustering consistent with the E17 measurements~(while also reproducing the clustering of simulated AGNs as predicted by MBII). We then discuss the implications of our model for the AGN population at $0.6\lesssim z\lesssim 2$, and make predictions for future measurements for the ongoing eBOSS survey.

Section \ref{methods_sec} summarizes the methods used, particularly the simulation and CLF formalism. Section \ref{AGN_properties_sec} discusses the scaling relations between the AGN properties as well as the properties of binary quasar pairs in MBII. In Sections \ref{MB2_CLF_sec} and \ref{LMF_sec}, we analyse the CLFs of AGNs in MBII, and build CLF models to probe the one-halo clustering. Section \ref{one_halo_sec} presents the one-halo clustering predictions by the CLF models. Section \ref{redshift_evolution_sec} discusses the redshift evolution of the AGN population as predicted by the CLF models and Section \ref{forecasts_sec} presents forecasts for the ongoing eBOSS-CORE survey. We summarize our results and make concluding remarks in Sections \ref{results_sec} and \ref{remarks_sec} respectively. Following E17, we present our results primarily in proper co-ordinates unless otherwise stated. Accordingly, `kiloparsecs' in proper co-ordinates shall be denoted by `kpc', and comoving coordinates shall be denoted by `ckpc'.     

\section{Methods}
\label{methods_sec}
\subsection{MassiveBlackII simulation}
\texttt{MBII} is a high-resolution cosmological hydrodynamic simulation which runs from $z=159$ to $z=0.06$. The simulation has a boxsize of $100~\mathrm{cMpc/h}$ and $2\times1792^3$ particles. The simulation used the cosmological parameters inferred from WMAP7 \citep{2011ApJS..192...18K} i.e. $\Omega_0=0.275$, $\Omega_l=0.725$, $\Omega_b=0.046$, $\sigma_8=0.816$, $h = 0.701$, $n_s=0.968$. The dark matter and gas particle masses are $1.1\times 10^7~M_{\odot}/h$ and $2.2\times 10^6~M_{\odot}/h$ respectively. The simulation was run using \texttt{P-GADGET}, an upgraded version of \texttt{GADGET} \citep{2005MNRAS.364.1105S}. In addition to the N-body gravity solver for the dark matter component and Smooth Particle Hydrodynamics (SPH) solver for the gas component, MBII incorporates subgrid physics modeling such as star formation \citep{2003MNRAS.339..289S}, black hole growth and associated feedback. Haloes and subhaloes were identified using the Friends-of-Friends (FOF) group finder \citep{1985ApJ...292..371D} and \texttt{SUBFIND} \citep{2005MNRAS.364.1105S} respectively. For more details, we refer the reader to \cite{2015MNRAS.450.1349K}.

\subsubsection{Black hole growth and associated feedback}
The prescription for black hole growth used in the simulation is adopted from \cite{2005Natur.433..604D} and \cite{2005MNRAS.361..776S}. A seed black hole of mass $5\times 10^{5}~M_{\odot}/h$ is placed in a halo of mass $\gtrsim 5\times 10^{10}~M_{\odot}/h$ (if the halo does not already contain a black hole). Once seeded, the black hole grows at a rate given by $\dot{M}_{bh}=\frac{4\pi G^2 M_{bh}^2 \rho}{(c_s^2+v_{bh}^2)^{3/2}}$; $\rho$ and $c_s$ are the density and sound speed of the cold phase of the ISM gas, and $v_{vh}$ is the relative velocity of the black hole w.r.t gas. The bolometric luminosity of the black halo is given by $\epsilon_r mc^2$ where $\epsilon_r$ is the radiative efficiency taken to be 0.1. $5\%$ of the energy released is thermodynamically (and isotropically) coupled to the surrounding gas \citep{2005Natur.433..604D}. Black holes can also grow via merging; two black holes are considered to be merged if they come within the spatial resolution of the simulation~(the SPH smoothening length) with a relative speed smaller than the local sound speed of the medium. For further details on the modeling of black hole growth, we refer readers to \cite{2012ApJ...745L..29D}.

\subsubsection{Identifying AGNs: Centrals and Satellites}
\label{central_satellite_identification}
Simulated AGNs are identified to be individual \textit{active} black holes accreting gas from the surrounding medium. Black holes are referred to as \textit{active}~(AGNs) if they radiate with a  bolometric luminosity $L_{\mathrm{bol}}\gtrsim 3\times10^{41} \mathrm{ergs/sec}$ and masses $M_{bh}\gtrsim 10^{6}~M_{\odot}/h$ (5 times the seed mass of the black hole). Within a host halo (FOF group), the most massive black hole is defined to be the \textit{central} AGN. Any other AGN within the same halo is defined to be a \textit{satellite} AGN.       

\subsection{AGN Clustering}
\label{method_clustering}
In this work, we focus on the two-point~(pairwise) clustering statistics quantified by the two-point correlation function. The one-component spatial correlation function is defined to be \begin{equation}
dN=n_{\mathrm{AGN}}(1+\xi(r)) 4\pi r^2 dr 
\end{equation}
where $r$ is the inter-particle distance; assuming a particle located at $r=0$, $dN$ is the total number of paired particles within a spherical shell at a distance $r$, thickness $dr$; $n_{\mathrm{AGN}}$ is the volume density of AGNs. Similarly, the two-component redshift space correlation function  is defined as
\begin{equation}
dN=n_{\mathrm{AGN}}(1+\xi(r_p,s)) 4\pi r_p dr_p ds 
\end{equation}
where $r_p$ and $s$ are the distance parallel and perpendicular (in redshift space) to the observer's line of sight, respectively; assuming a particle located at the origin i.e. $(r_p=0,s=0)$, $dN$ is the total number of paired particles within a cylindrical shell at projected distance~(inner radius) $r_p$, thickness $dr_p$ and height $ds$.

The projected correlation function $w_p(r_p)$ is defined as 
\begin{equation}
w_p(r_p)=\int_{0}^{\infty}\xi(r_p,s) ds=2\int_{r_p}^{\infty}\frac{r\xi(r)}{\sqrt{r^2-r_p^2}} dr.
\label{projection}
\end{equation}
Recent works \citep{2006AJ....131....1H,2012MNRAS.424.1363K,2017MNRAS.468...77E} on measurements of quasar clustering at small ($\sim \mathrm{kpc}$) scales have used the ``Volume averaged projected correlation function" $\bar{W}_p$. For quasar pairs within transverse separations of $r_{\mathrm{min}}<r_p<r_{\mathrm{max}}$ and a line of sight velocity separation of $v_{\mathrm{max}}$, this statistic is defined as
\begin{eqnarray}
   \nonumber \bar{W}_p(z,r_{\mathrm{min}},r_{\mathrm{max}})=\frac{1}{V_{\mathrm{shell}}} \int_{-v_{\mathrm{max}}/aH(z)}^{v_{\mathrm{max}}/aH(z)}\int_{r_{\mathrm{min}}}^{r_{\mathrm{max}}}\xi(r_p,s,z)\\ \times 2 \pi r_p d r_p ds
    \label{W_p}
\end{eqnarray}
where $V_{\mathrm{shell}}=\pi(r_{\mathrm{max}}^{2}-r_{\mathrm{min}}^{2}) \frac{2v_{\mathrm{max}}}{aH(z)}$. Unlike $w_p$ which has a dimension of length, $\bar{W}_p$ is dimensionless and is essentially the two component correlation function $\xi(r_p,s)$ averaged over the volume $V_{\mathrm{shell}}$. In practice, if the velocity separation is large enough (true in our case with $v_{\mathrm{max}}=2000$ km/sec), the redshift space distortions can be effectively removed by the integration over $s$ in Eq.~(\ref{W_p}), and the following approximation can be used to convert between $w_p$ and $\mathrm{W}_p$
\begin{equation}
    w_p\approx2\frac{v_{\mathrm{max}}}{aH(z)}\bar{W}_p;
    \label{conversion}
\end{equation}
(see also Eq.~A2 of \citealt{2012ApJ...755...30R}).

\subsection{Conditional luminosity function modeling}
\label{method_CLF}
The Conditional Luminosity Function (CLF) approach has been widely used for clustering analyses of galaxies \citep[and references therein]{2006MNRAS.365..842C,2017MNRAS.471.2022T}. In this work, we shall be constructing a CLF model for the one-halo clustering of AGNs. The CLF, denoted by $\Phi(L,M_h)$, measures the distribution of AGN bolometric luminosities $L$ within a halo of mass $M_h$. Its definitive properties are 1)
\begin{equation}
\int~\Phi(L,M_h)~\frac{dn}{dM_h}~dM_h=\Phi(L)  
\end{equation}
where $\Phi(L)$ is the overall luminosity function, and 2)
\begin{equation}
\int_{L_{\mathrm{min}}}^{\infty}~\Phi(L',M_h)~d\log_{10}L'=\left<N(L>L_{\mathrm{min}},M_h)\right> 
\label{CLF_to_HOD}
\end{equation}
where $\left<N(L>L_{\mathrm{min}},M_h)\right>$ is the mean halo occupation number of AGNs with $L>L_{\mathrm{min}}$. Note that $L$ is written in units of $\mathrm{ergs/sec}$ throughout the paper unless stated otherwise.  
We separate our CLF into contributions from central and satellite AGNs (according to Section \ref{central_satellite_identification})
\begin{equation}
    \Phi(L,M_h)=\Phi_{\mathrm{cen}}(L,M_h)+\Phi_{\mathrm{sat}}(L,M_h).
\end{equation}
As we shall see in Section \ref{MB2_CLF_sec}, CLFs for central AGNs can be modelled as a log-normal distribution 
\begin{eqnarray}
\nonumber \Phi_{\mathrm{cen}}(L,M_h)=f_{\mathrm{active}}~ P_{\mathrm{cen}}(L,M_h)\\= f_{\mathrm{active}}~\frac{1}{2 \pi \sigma_{\mathrm{cen}}} \exp{\left(-\frac{(\log_{10}{L}-\log_{10}{L^*_{\mathrm{cen}}})^2}{\sigma_{\mathrm{cen}}^{2}}\right)} ,
\label{central_CLF_eqn}
\end{eqnarray}
where $P_{\mathrm{cen}}$ is the normalized probability distribution of luminosities of central AGNs within a halo of mass $M_h$. $L^*_{\mathrm{cen}}$ is the average central AGN luminosity, and $\sigma_{\mathrm{cen}}$ is the width of the log-normal distribution. $f_{\mathrm{active}}$ is the normalization of the central AGN CLF, and refers to the fraction of haloes which host at least one \textit{active} AGN. CLFs for satellite AGNs can be modelled as a Schechter distribution
\begin{eqnarray}
\nonumber \Phi_{\mathrm{sat}}(L,M_h)=\left<N_{\mathrm{sat}}\right>P_{\mathrm{sat}}(L,M_h)\\ = Q_{\mathrm{sat}} \left(\frac{L}{L^*_{\mathrm{sat}}}\right)^{\alpha_{\mathrm{sat}}}\exp{\left(-\frac{L}{L^*_{\mathrm{sat}}}\right)}
\label{satellite_CLF_eqn}
\end{eqnarray}
where $\left<N_{\mathrm{sat}}\right>$ and $P_{\mathrm{sat}}$ are, respectively, the average number and the probability distribution of luminosities of satellite AGNs within a halo of mass $M_h$. $L^{*}_{\mathrm{sat}}$ is a measure of the most luminous satellite for a given halo of mass $M_h$, $\alpha_{\mathrm{sat}}$ determines the slope of the distribution for $ L<L^{*}_{\mathrm{sat}}$, and $Q_{\mathrm{sat}}$ is the overall normalization.

$\xi(r)$ can be decomposed as
\begin{equation}
    \xi(r)=\xi_{1h}(r)+\xi_{2h}(r)
    \label{correlation}
\end{equation}
where $\xi_{1h}$ and $\xi_{2h}$ are `one-halo' and `two-halo' contributions respectively. The one-halo contribution for the power spectrum $P_{1h}$ is given by
\begin{eqnarray}
\nonumber P_{1h}(k)=\frac{1}{n_{\mathrm{AGN}}^2}\int(\left<N_{\mathrm{cen}}N_{\mathrm{sat}}\right>+\left<N_{\mathrm{sat}}(N_{\mathrm{sat}}-1)\right>) \frac{dn}{dM_h}\\ u(k,M_h)^2 dM_h 
\end{eqnarray}
where $\left<N_{\mathrm{cen}}N_{\mathrm{sat}}\right>$ and $\left<N_{\mathrm{sat}}(N_{\mathrm{sat}}-1)\right>$ are the expected number of central-satellite and satellite-satellite pairs, $u(k,M_h)$ is the Fourier transform of the normalized satellite AGN density profile, and $\frac{dn}{dM_h}$ is the halo mass function. We assume that central and satellite occupations are independent i.e. $\left<N_{\mathrm{cen}}N_{\mathrm{sat}}\right>=\left<N_{\mathrm{cen}}\right>\left<N_{\mathrm{sat}}\right>$, and satellite distributions are Poisson i.e. $\left<N_{\mathrm{sat}}(N_{\mathrm{sat}}-1)\right>=\left<N_{\mathrm{sat}}\right>^2$. $\left<N_{\mathrm{cen}}\right>$ and $\left<N_{\mathrm{sat}}\right>$ can be obtained from $\Phi_{\mathrm{cen}} (L,M_h)$ and $\Phi_{\mathrm{sat}} (L,M_h)$ respectively using Eq.~(\ref{CLF_to_HOD}). $\xi_{1h}(r)$ can then be determined using 
\begin{equation}
    \xi_{1h}(r)=\frac{1}{2\pi^2}\int~dk~k^2 P_{1h}(k) \frac{\sin{kr}}{kr}  .
\end{equation}

The two-halo contribution is given by 
\begin{equation}
    \xi_{2h}(r)=\xi_{\mathrm{matter}}(r)~b_{\mathrm{eff}}^2
\end{equation}
where $\xi_{\mathrm{matter}}(r)$ is the matter power spectrum. Here, $b_{\mathrm{eff}}$ is the effective bias of AGNs given by 
\begin{equation}
    b_{\mathrm{eff}}=\frac{1}{n_{AGN}}\int dM_h (\left<N_{\mathrm{cen}}\right>+\left<N_{\mathrm{sat}}\right>)~\frac{dn}{dM_h} b(M_h)
    \label{bias}
\end{equation}
where $b(M_h)$ is the halo bias.

We make the following additional assumptions in our modeling:
\begin{itemize}
\item We use the halo mass function from \cite{2008ApJ...688..709T}. MBII mass functions are consistent with \cite{2008ApJ...688..709T} as shown in \cite{2015MNRAS.450.1349K}. 

\item For the two-halo term, we use the linear halo bias model from \cite{2010ApJ...724..878T} and the non-linear matter power spectrum predicted by MBII. 

\item For the one-halo term, we use a power-law satellite profile with exponent `-2', which is consistent with the radial distribution of satellite AGNs in MBII, as discussed in Appendix \ref{satellite_profile_sec}.
\end{itemize}

Our objective is to use the methods summarized in Section \ref{method_clustering} and \ref{method_CLF} to obtain theoretical predictions for $\bar{W}_p$ and compare with observed measurements. For the sample of simulated AGNs (assuming it is large enough), one can use Eq.~(\ref{Wp_from_simulation}) to obtain $\bar{W}_p$. To determine $\bar{W}_p$ using CLF modeling, we need to first obtain $\xi(r)$ using Eqs.~(\ref{correlation})-(\ref{bias}), we can then use Eqs.~(\ref{projection}) and (\ref{conversion}) to obtain the predictions for $w_p$ and $\bar{W}_p$.
\subsection{Determining $W_p$}
\label{determine_wp}
One of our primary goals in this work is to determine $W_p$ and compare with observational constraints; we have used two methods to do so:
\begin{itemize}
\item \textit{Method 1~(using simulated objects): For a given snapshot, if the number of objects is large enough we can use 
\begin{equation}
    \bar{W}_p=QQ/\left<QR\right>-1
    \label{Wp_from_simulation}
\end{equation}
where $QQ$ is the number of AGN-AGN pairs within a projected distance of $r_{\mathrm{min}}$ to $r_{\mathrm{max}}$ and a line of sight separation of $v_{\mathrm{max}}/aH(z)$ in redshift space; $\left<QR\right>$ is the expected number of AGN-random pairs inside the same cylindrical shell (the random objects are uniformly distributed over the volume).} 

\item \textit{Method~2~(using CLF modeling)}: $\xi(r)$ can be determined from the CLFs using the methods outlined in Section \ref{method_CLF}. $W_p$ can then be determined from $w_p$ by using Eq.~(\ref{conversion}), where $w_p$ can be determined from $\xi(r)$ using Eq.~(\ref{projection}).
\end{itemize}

\section{AGN properties in MBII}
\label{AGN_properties_sec}
\subsection{Scaling relations}
\label{scaling_relations_sec}
\begin{figure*}
\includegraphics[width=\textwidth]{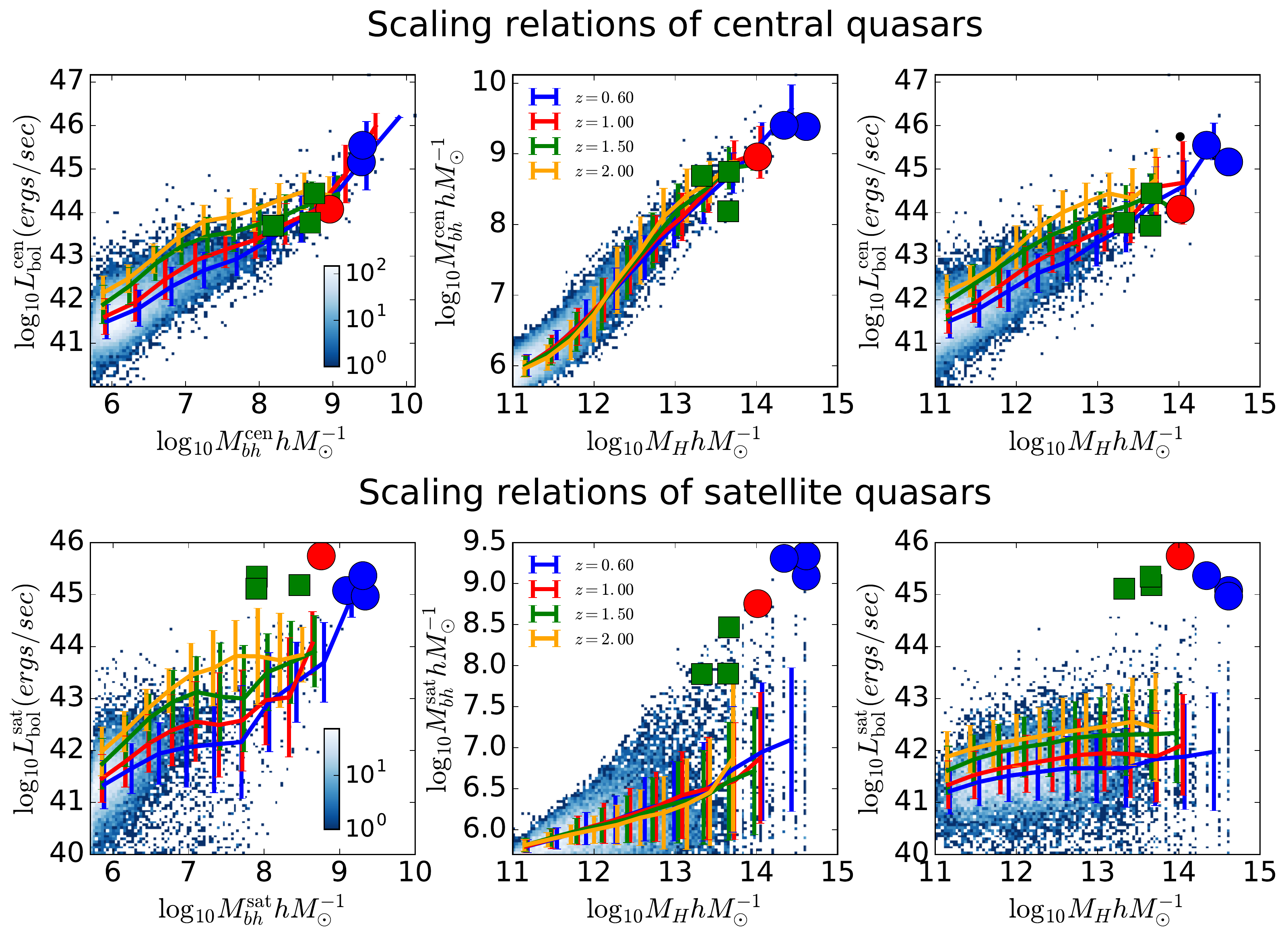}
\caption{Scaling relations of central (top panels) and satellite (bottom panels) quasars. The blue histogram shows the relations for $z=0.6$. Solid lines show the mean relations at various redshifts and the error bars show $1\sigma$ scatter. Filled circles in all the panels correspond to the central-satellite pairs with $g<20.85$ (the E17 magnitude limit). There are 3 such pairs at $z=0.6$ (blue circles) and 1 pair at $z=1$ (red circle) (Note that for the blue circles, two of the satellites are paired with the same central, as shown in the leftmost panel of Figure 2). The green squares show the central-satellite pairs above the eBOSS limit~($g<22$) at a target redshift of $z=1.5$.}
\label{scaling_relations_fig}
\end{figure*}

\begin{figure*}
\addtolength{\tabcolsep}{-20pt}
\begin{tabular}{ccc}
\includegraphics[width=7cm]{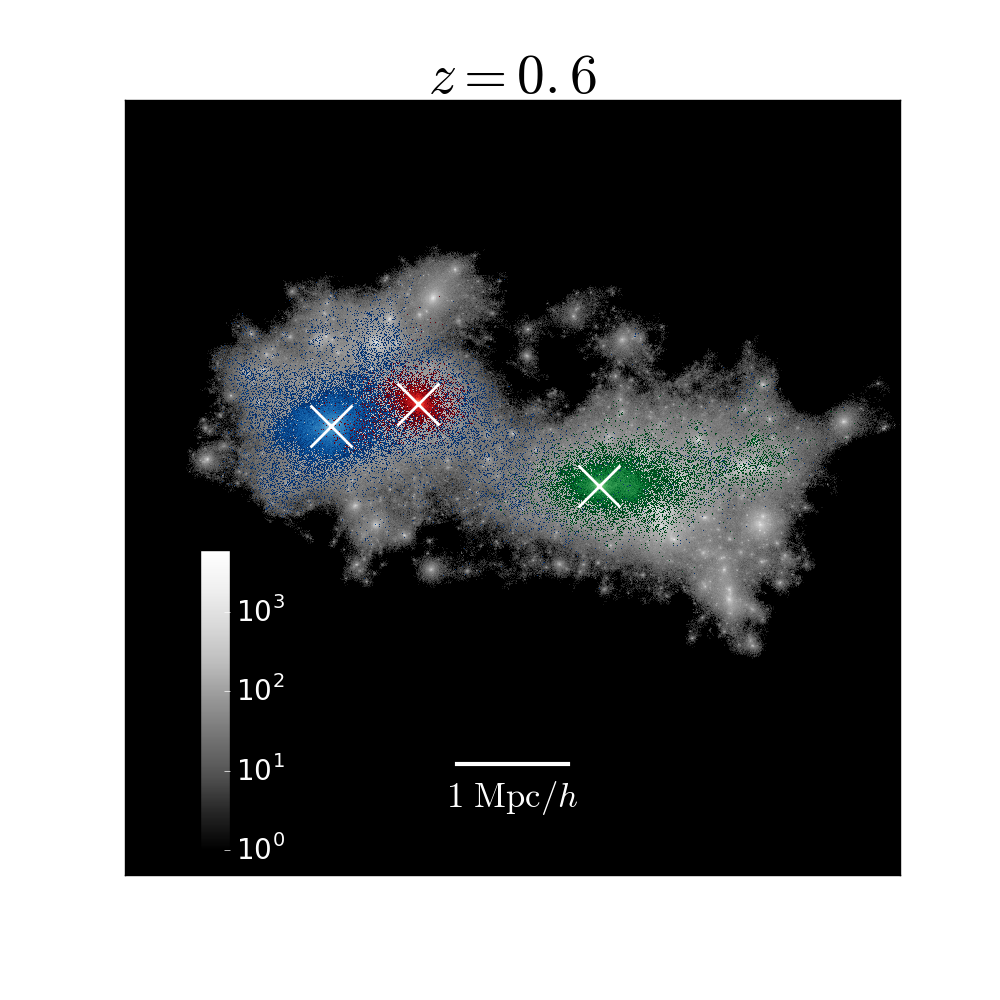}&
\includegraphics[width=7cm]{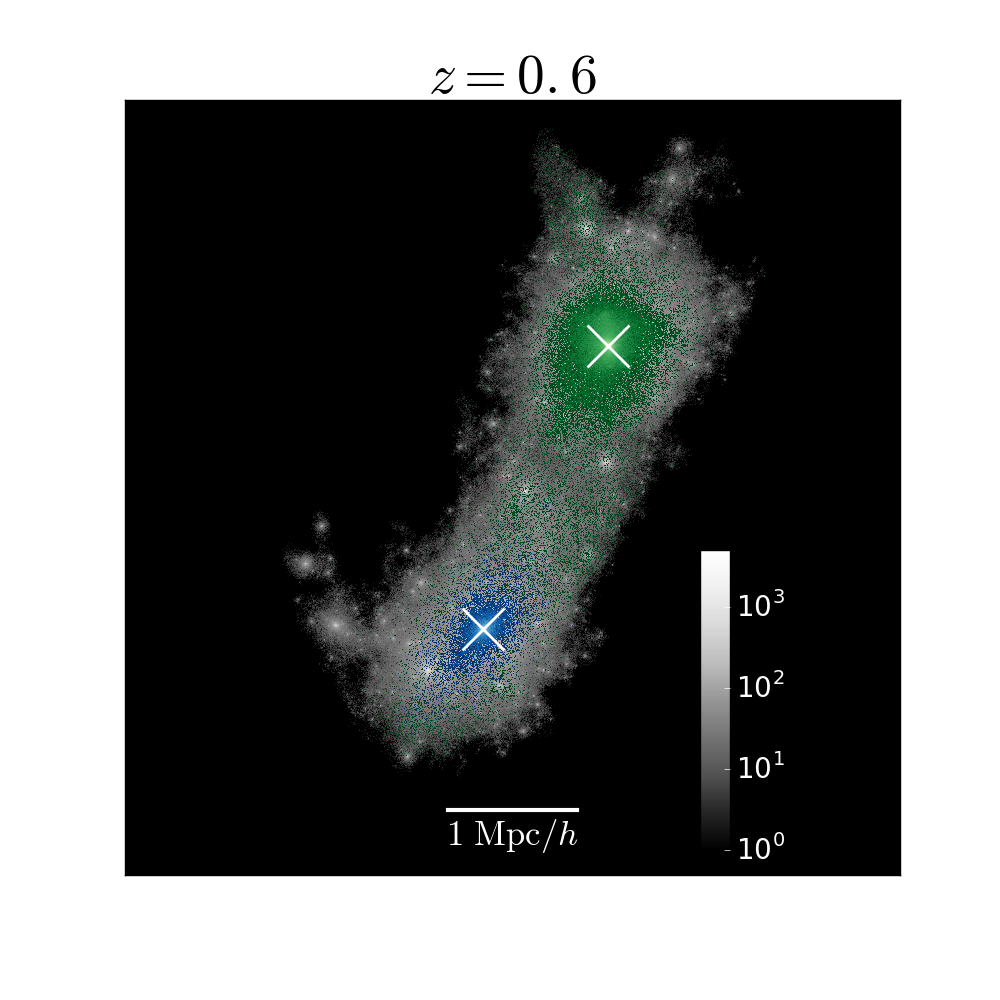}&
\includegraphics[width=7cm]{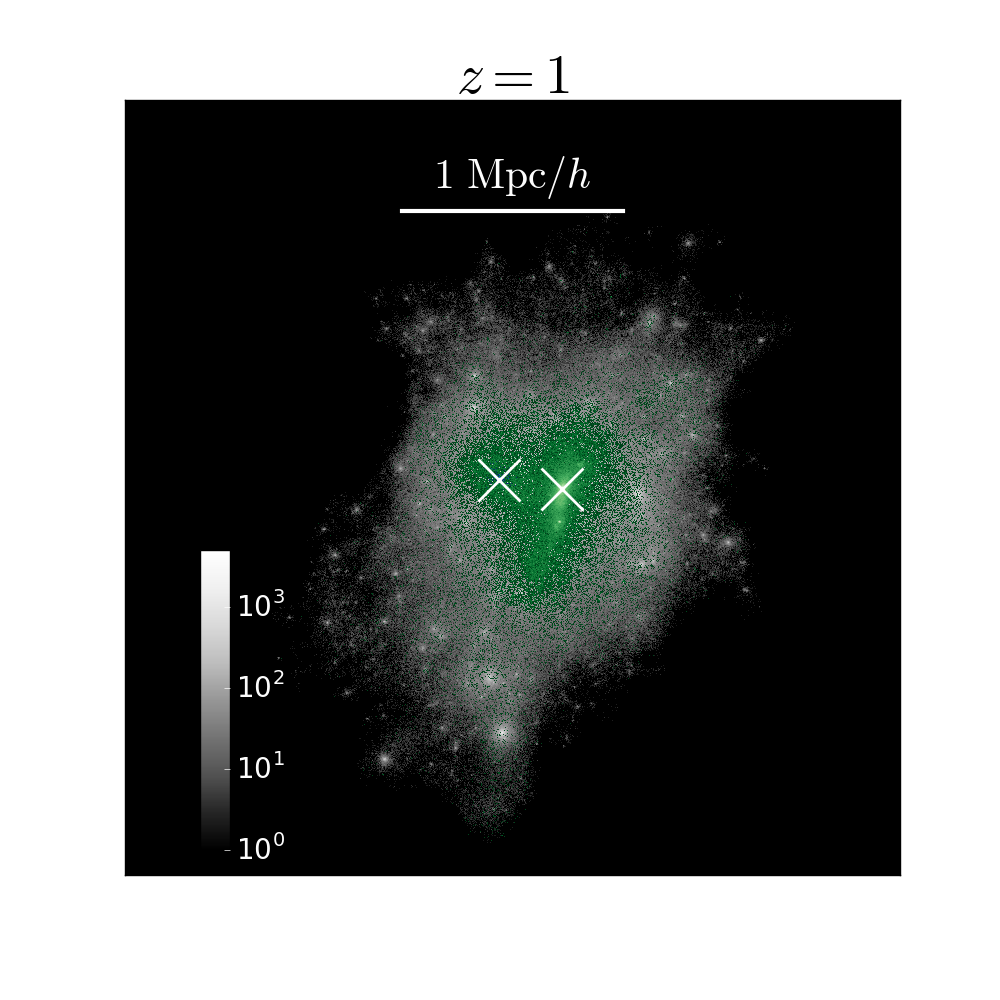}
\end{tabular}
\caption{Illustrations of quasar pairs
with $g<20.85$ at $z=0.6$ snapshot
(white crosses) in MBII. The stars in their host galaxies are shown in colors. The green histogram shows the host galaxy of the central (most massive black hole) quasar, and the red and blue histograms show the host galaxies of the satellite quasars. The grey histogram shows the host dark matter halo (FOF group), for the pairs indicated by filled circles in Figure \ref{scaling_relations_fig}. The stellar masses of the host galaxies are $2\times10^{12},~ 1\times10^{12},~ 1\times10^{12}~M_{\odot}$ for the crosses within the green, blue and red histograms respectively. Properties of the quasars and their hosts are shown in Figure~\ref{growth_history}.}

\label{illustrations}
\end{figure*}

\begin{figure}
\includegraphics[width=8cm]{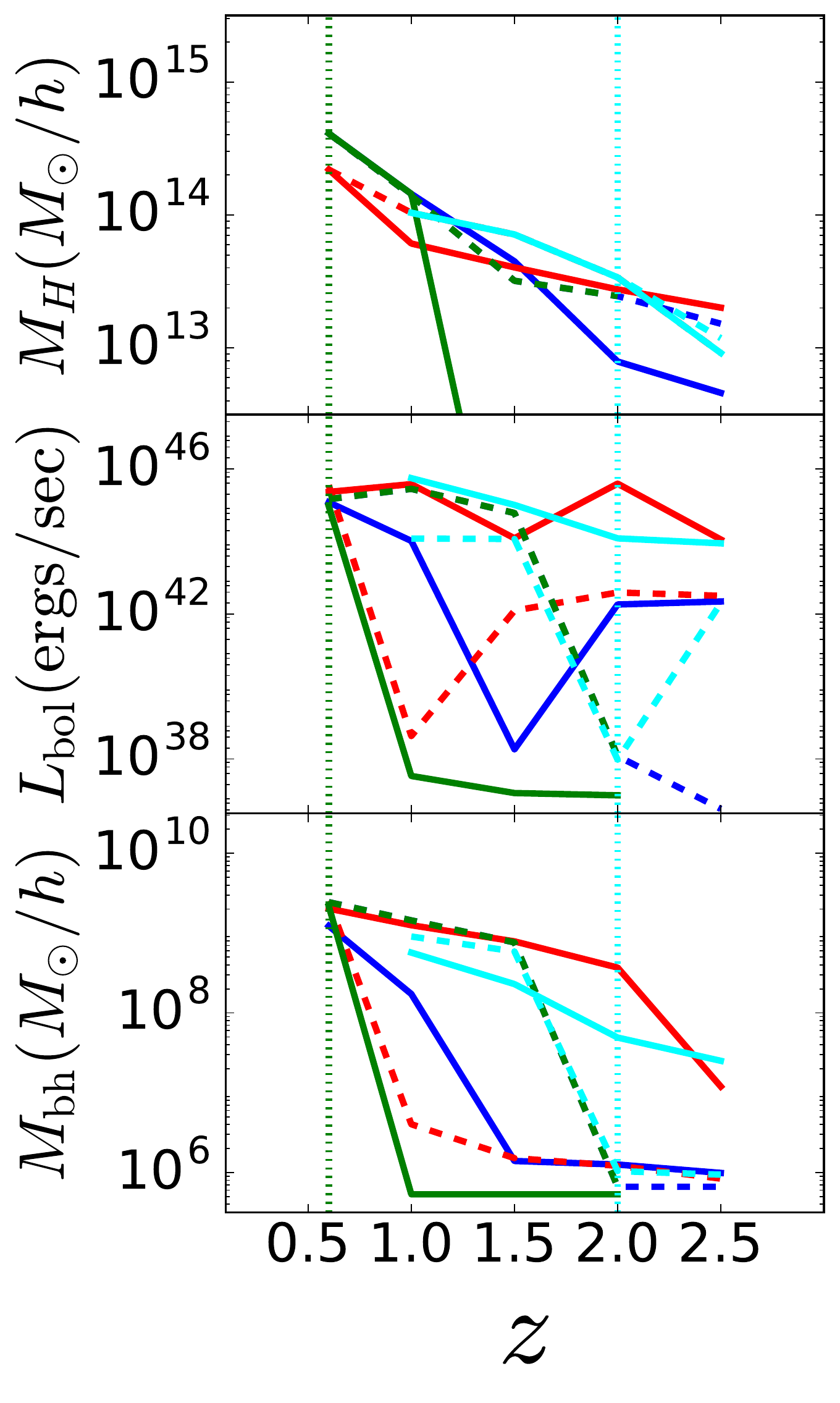}
\caption{The blue, red and green lines show the redshift evolution of the properties of three simulated quasar pairs at $z=0.6$. The solid and dashed lines correspond to the satellite and central quasars respectively. The \textbf{top panel} shows the evolution of the host halo masses and the vertical lines correspond to the simulation snapshot when the host haloes merge. The \textbf{middle panel} shows the quasar bolometric luminosities. The \textbf{bottom panel} shows the black hole masses. The cyan lines correspond to a pair at $z=1$ where the satellite~(but not the central) AGN becomes brighter than $g=20.85$ at $z=1$ after the halo merger at $z=2$.}
\label{growth_history}
\end{figure}

There are a total of 57046 black holes at $z=2$, of which 49299 are \textit{active} and 7747 are \textit{inactive}~(as defined in Section \ref{central_satellite_identification}). By $z=0.6$ there are a total of 76895 black holes, of which 28850 are \textit{active} and 48045 are \textit{inactive}. We first present the scaling relations between various properties of the full AGN population (drawn from this sample of black holes) in the MBII simulation. Figure \ref{scaling_relations_fig} presents the relations between the bolometric luminosity $L_{\mathrm{bol}}$, black hole mass $M_{\mathrm{bh}}$ and host halo mass~(mass of the FOF group) $M_h$ of the quasars in MBII for central BHs (residing in the most massive (central) galaxies) and satellite BH/AGN within the host halo (which are typically associated with a satellite galaxy). The histograms show the full scatter at $z=0.6$. The solid lines from blue to orange show the redshift evolution (the mean at at each redshift) from $z=0.6$ to $z=2$. As expected, we find that the black hole mass is correlated with the bolometric luminosity for both central and satellite AGNs (see leftmost panels). Observational estimates of black hole masses of SDSS quasars \citep{2013BASI...41...61S} do find evidence of such a correlation. Also, for a given black hole mass, bolometric luminosity increases with increasing redshift for both central and satellite AGNs. $M_{bh}$ vs. $M_h$ relations are shown in the middle panels; there is no significant redshift evolution in the $M_{bh}$ vs. $M_h$ relation. The $M_{bh}$-$L_{bol}$ and $M_{bh}$-$M_h$ relations then inevitably produce the positive correlation between $L_{bol}$ vs. $M_h$ that is shown in the rightmost panels of Figure \ref{scaling_relations_fig}. Note that for satellite black holes the slope of $L_{bol}$ vs. $M_h$ is significantly smaller than for central AGNs. In a nutshell, we find that the relation between AGN luminosity and halo mass is primarily governed by the associated black hole mass - halo mass relation. As a result, more luminous AGNs live in more massive haloes. We also find that the correlation is stronger for central AGNs compared to satellite AGNs. Furthermore, AGNs for fixed halo mass become more luminous with increasing $z$.
\subsection{Quasar pairs in MassiveBlackII}
\label{pairs_in_MBII}
The primary objective of our work is to interpret the small-scale clustering of quasar pairs measured in E17. These pairs are limited to a magnitude of $g<20.85$. Within the MBII simulation volume, we find only 3 pairs with $g<20.85$ at $z=0.6$ at distances of $\sim1-4~\mathrm{Mpc}$.; there are no such pairs at $z=1,1.5,2$. These numbers are reasonably consistent with the observed luminosity functions of quasars; as discussed later in Section \ref{LMF_sec}. The properties of these quasar pairs are marked as filled blue circles in Figure \ref{scaling_relations_fig}, which shows that the central quasars lie reasonably close to the mean trends (solid lines).  However, for the satellite quasars, we find (see Figure \ref{scaling_relations_fig}) that both black hole masses ($M_{bh}\sim10^{9}~M_{\odot}/h$ and luminosities ($L_{\mathrm{bol}}\sim 10^{45}~\mathrm{ergs/sec}$) are $\sim10^2-10^3$ times higher than the typical values ($M_{bh}\sim10^{7}~M_{\odot}/h$ and $L_{\mathrm{bol}}\sim3 \times 10^{41}~\mathrm{ergs/sec}$). In fact, the black hole masses and luminosities of these satellite AGNs are comparable to those of central AGNs, making them an extreme and rare subset of the satellite population. This hints at the possibility of these pairs originating from recent halo mergers, such that both members of the pairs were two central AGNs prior to the merger.

In order to further investigate whether these pairs are activated by mergers, we look into the properties of their host galaxies and haloes, and their progenitors. Figure \ref{illustrations} (left and middle panels) shows the projected positions of 3 pairs (at $z=0.6$) with $g<20.85$ along with their host galaxies and haloes. The host haloes are shown as grey histograms and have masses $\sim 1-3 \times 10^{14}~M_{\odot}/h$. The morphology of the host haloes in both cases (left and middle panels) suggest that they are involved in a major merger. The colored histograms show the host galaxies of these individual quasars. The green histograms show the host galaxy of the central AGN; red and blue histograms show the host galaxy of the satellite AGNs. Note that host galaxies of the central AGNs also consistently have slightly higher (by factors of 1.5-2.5) stellar masses compared to satellites, making them consistent with the usual definition of ``central galaxies". As we can see, these simulated quasar pairs reside in separate (central and satellite) galaxies within the same halo. They all have roughly comparable (within a half order of magnitude) stellar masses ranging from $M_*\sim 1-3\times 10^{12}~M_{\odot}/h$, and are both among the most massive galaxies in the simulation. This suggests that they were initially two centrals in different haloes, which then merged to form a central-satellite pair (according to our definition in Section \ref{central_satellite_identification}) with similar stellar masses. 
We now look at the growth history of the three $g<20.85$ quasar pairs and their host haloes. The panels in Figure \ref{growth_history} show the evolution of $M_h$, $L_{\mathrm{bol}}$ and $M_{\mathrm{bh}}$. The pairs at $z=0.6$ are denoted by red, blue and green lines. The host halo masses of their progenitor AGNs (top panels) are different at $z=1$, implying that a merger happened between $z=1$ and $z=0.6$. If we look at the evolution of $L_{\mathrm{bol}}$ and $M_{\mathrm{bh}}$ in the middle and bottom panels, we see that the activity of one or both members is significantly enhanced after the merger. Particularly for the solid  green and dashed red lines, $L_{\mathrm{bol}}$ increases from $\sim10^{38}~\mathrm{ergs/sec}$ at $z=1$ to $\sim 10^{44.5}~\mathrm{ergs/sec}$; consequently, $M_{\mathrm{bh}}$ rises from $\sim10^6~M_{\odot}/h$ at $z=1$ to $\sim10^9~M_{\odot}/h$ at $z=0.6$. This confirms that the AGN activity required for the formation of these bright quasar pairs was indeed triggered by recent halo mergers. 

The cyan lines correspond to the pair at $z=1$ (shown as red circles in Figure \ref{scaling_relations_fig}), and is also an interesting example demonstrating the extent to which halo mergers trigger AGN activity (in this pair, only the satellite AGN is bright enough to be observable) in the simulation. At z=2, the mass of the central AGN (dashed cyan line in the bottom panel) progenitor is 2 orders of magnitude less than that of the satellite (solid cyan line in the bottom panel); but the halo merger around $z\sim2$ triggered AGN activity in the central such that it surpasses the black hole mass of the satellite AGN at $z=0.6$. Similar to $g<20.85$ pairs, we also report halo merger-driven AGN activity for $z\sim1.5$ satellites above the magnitude limit of eBOSS-CORE quasars \citep[$g\lesssim22$;][]{2015ApJS..221...27M}, which are shown as green squares in Figure \ref{scaling_relations_fig} (bottom panels).        
It is not possible to directly calculate the (statistical) small-scale clustering of quasars with so few simulated objects. In the next section, we therefore investigate the conditional luminosity functions (CLFs) of MBII AGNs; we shall then build a CLF model to study the small-scale clustering of $g<20.85$ quasars.     

\section{Conditional Luminosity Functions (CLFs) of MassiveBlackII AGNs}
\label{MB2_CLF_sec}
\begin{figure*}
\includegraphics[width=\textwidth]{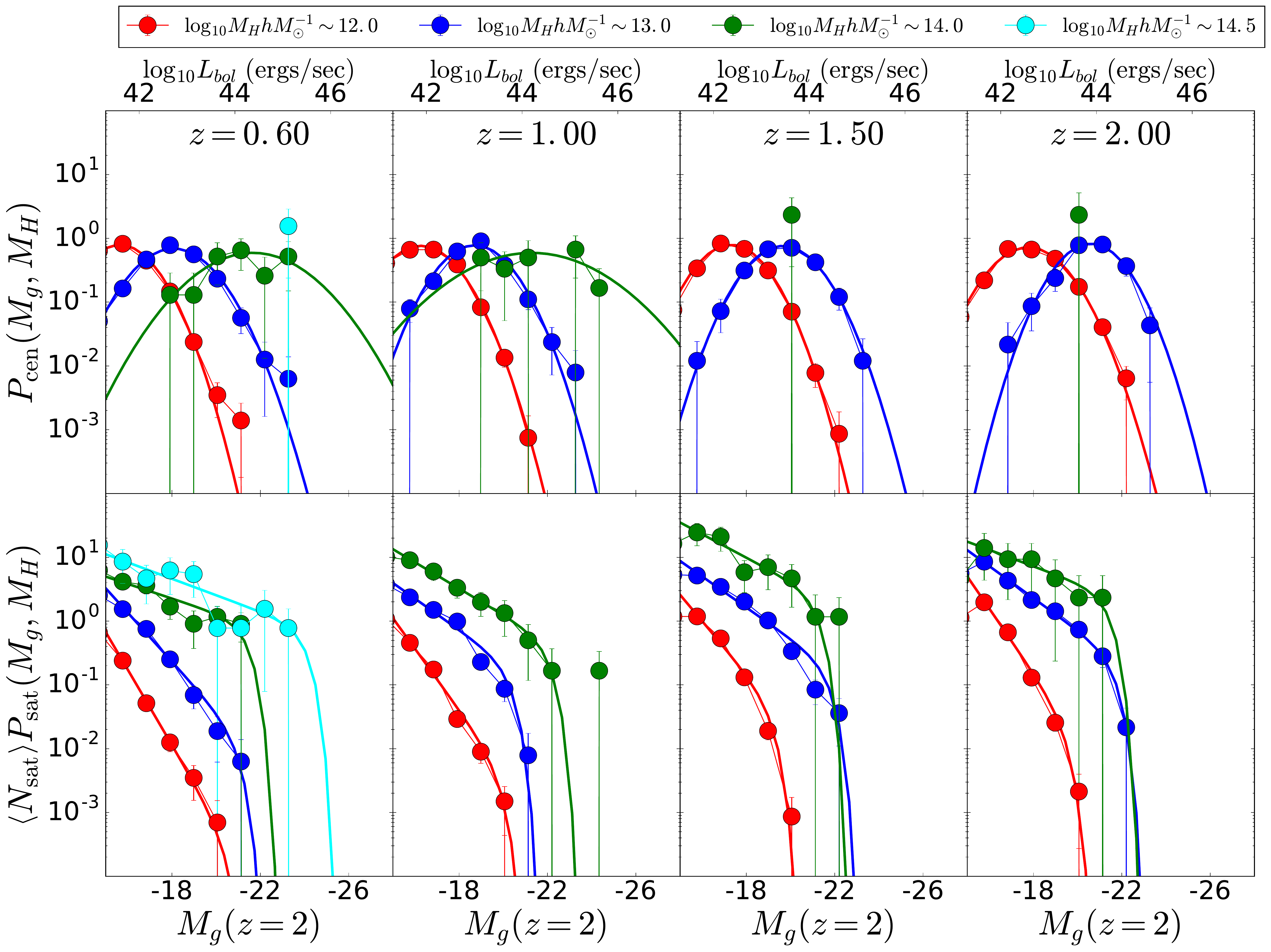}
\caption{Filled circles show Conditional luminosity functions (CLFs) of AGNs in MBII. \textbf{Top Panels} correspond to CLFs of centrals AGNs and solid lines show the best fit log-normal distributions (Eq.~\ref{central_CLF_eqn}). \textbf{Bottom Panels} correspond to CLFs of satellite AGNs and solid lines show the best fit Schechter distribution (Eq.~\ref{satellite_CLF_eqn}).} 
\label{CLF_fig}
\end{figure*}
\label{CLF_sec}
\begin{figure*}
\includegraphics[width=\textwidth]{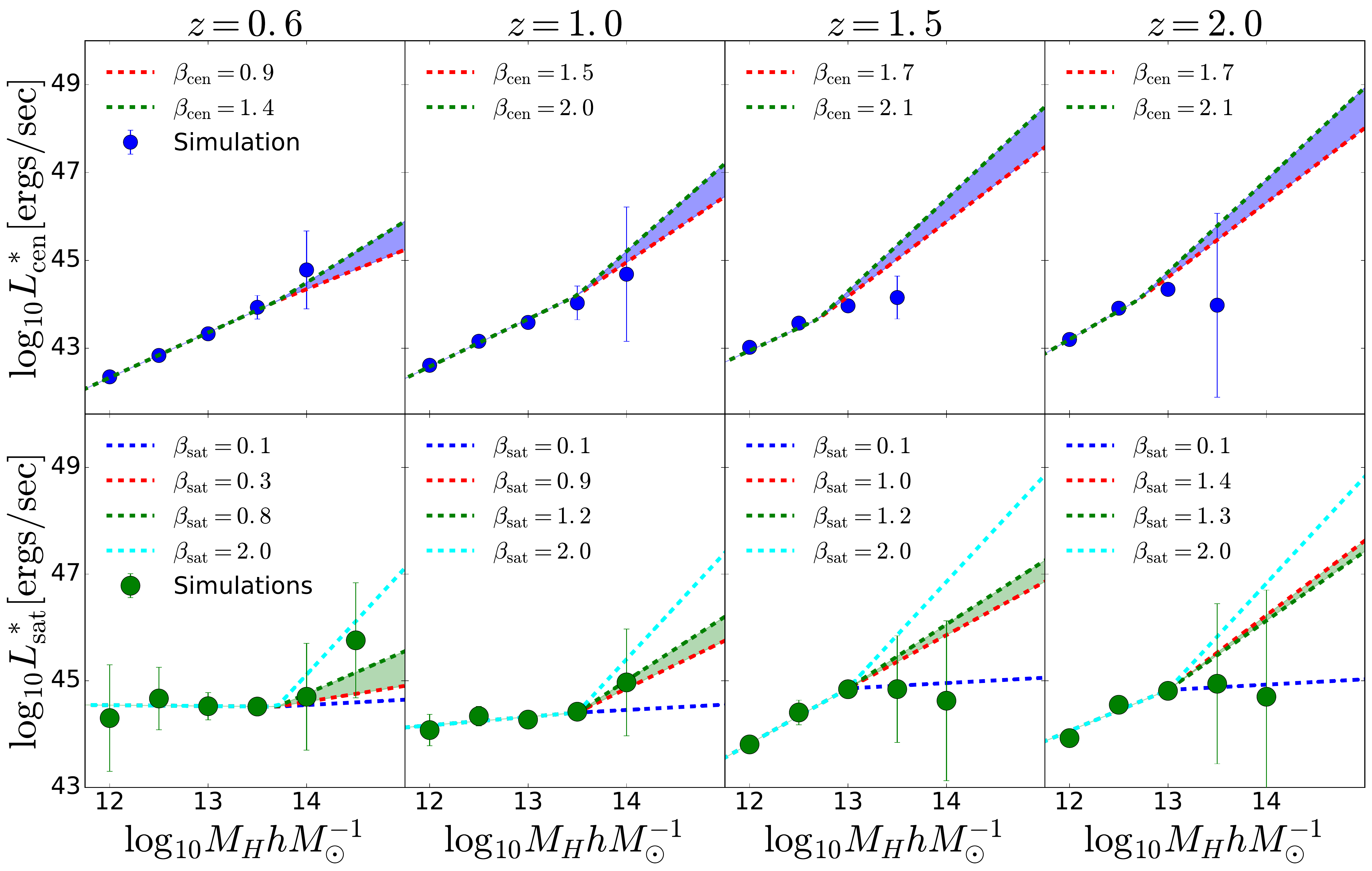}
\caption{\textbf{Top panels:} The blue filled circles show the best fitting values of $L^{*}_{\mathrm{cen}}$ (for central AGNs) at $M_h\lesssim 10^{14} M_{\odot}/h$; errorbars correspond to $1\sigma$ errors for the parameter estimates (determined from the covariance matrix returned by \texttt{scipy.optimize.curve_fit}). The red and blue dashed lines show the range of $\beta_{\mathrm{cen}}$ values allowed by the observed luminosity functions shown in Figure \ref{LMF_sims_vs_obs_fig} (as described in the legend). \textbf{Bottom panels:} The green filled circles show the best fitting values of $L^{*}_{\mathrm{sat}}$ (for satellite AGNs) at $M_h\lesssim 10^{14} M_{\odot}/h$; errorbars show the $1\sigma$ errors. The dashed lines show the various possible models of $\beta_{\mathrm{sat}}$ values for $M_h\gtrsim 10^{14} M_{\odot}/h$. The red and green dashed lines (enclosing the shaded green region) show the range of $\beta_{\mathrm{sat}}$ values (as described in the legend) inferred from the observed small-scale clustering measurements in E17 shown in Figure \ref{average_projected_clustering}.}
\label{Lstar_models_fig}
\end{figure*}
Filled circles in the top and bottom panels of Figure \ref{CLF_fig} show the MBII predictions for the Conditional Luminosity Functions (CLFs) for central and satellite AGNs respectively. We express the CLFs as a function of $g$-band absolute magnitude, $M_g~(z=2)$, which can be obtained from the bolometric luminosity $L_{\mathrm{bol}}$ using Eq.~(1) and Eq.~(2) of \cite{2009ApJ...697.1656S} and \cite{2009MNRAS.399.1755C} respectively,
\begin{eqnarray}
    M_i(z=2)=90-2.5\log_{10}\frac{L_{\mathrm{bol}}}{{\mathrm{ergs~sec^{-1}}}}\\
    M_g(z=2)=M_i(z=2)+2.5\alpha_{\nu}\log_{10}{\frac{4670~\mathrm{\AA}}{7471~\mathrm{\AA}}}
\end{eqnarray}
where $\alpha_{\nu}=-0.5$.

For the central AGNs, the CLFs can be well described by a log-normal distribution given by Eq.~(\ref{central_CLF_eqn}) (Note: we use the \texttt{scipy.optimize.curve_fit} package to perform our fits). A key parameter of interest is the overall normalization $f_{\mathrm{active}}$. In principle, $f_{\mathrm{active}}$ can lie anywhere between 0 and 1 because AGNs have finite lifetimes and not all haloes~(including the very massive $M_h\gtrsim10^{14}~M_{\odot}/h$ haloes) may necessarily host an \textit{active} black hole. However~(see Appendix \ref{black_hole_duty_cycles} for a detailed discussion), it turns out that all MBII haloes with $M_{h}\gtrsim10^{12}~M_{\odot}/h$ always host at least one \textit{active}~(central) black hole~(also see Figure \ref{scaling_relations_fig}: top right panel). In other words, the overall normalization ($f_{\mathrm{active}}$) of $\Phi_{\mathrm{cen}}$ is unity. While there is no reason \textit{a priori} for why $f_{\mathrm{active}}$ should be 1 given that quasars have finite lifetimes, this finding is also consistent with previous works on hydrodynamic simulations \citep{2012MNRAS.419.2657C,2017MNRAS.466.3331D}. \cite{2012ApJ...755...30R,2018MNRAS.477...45M} also implicitly assume $f_{\mathrm{active}}=1$ in its chosen parametrization of the mean halo occupation of central AGNs. The other key parameter of interest is $L^*_{\mathrm{cen}}$, which measures the characteristic mean luminosity of central AGNs as a function of halo mass. Filled blue circles in the top panel of Figure \ref{Lstar_models_fig} show the best fit values of $L^*_{\mathrm{cen}}$ for different halo mass bins. We can see that the maximum halo masses probed by MBII at different redshifts are $M_h\sim10^{14},10^{13.5},10^{13},10^{13}~M_{\odot}/h$ for $z\sim 0.6,1.0,1.5,2.0$ respectively. Within this range,  $L^*_{\mathrm{cen}}$ increases with halo mass $M_h$ as a power law; this is expected from the scaling relations shown in the top-right panel of Figure \ref{scaling_relations_fig}.

For the satellite AGNs, we find that CLFs can be well fit by a Schechter distribution which is given by Eq.~(\ref{satellite_CLF_eqn}). Here, the key parameter of interest is $L^*_{\mathrm{sat}}$, which corresponds to the ``edge" of the satellite CLF; i.e. the most luminous satellite quasars within haloes of a given mass $M_h$. $L^*_{\mathrm{sat}}$ therefore would be sensitive to halo mergers if mergers indeed triggered the formation of satellite quasars that are $10^3$ times more luminous than the average values (as seen with the simulated quasar pairs in Section \ref{pairs_in_MBII}). Filled green circles in the bottom panel of Figure \ref{Lstar_models_fig} show the best fit values of $L^*_{\mathrm{sat}}$ for different halo mass bins. As in the case of centrals, the maximum halo masses probed by MBII at different redshifts are $M_h\lesssim10^{14},10^{13.5},10^{13},10^{13}$ for $z\sim 0.6,1.0,1.5,2.0$ respectively. At $z=0.6,1.0$, we do not find a significant dependence of $L^*_{\mathrm{sat}}$ on $M_h$ for $M_h\lesssim10^{14},10^{13.5}~M_{\odot}/h$. However, at $z=1.5,2.0$ $L^*_{\mathrm{sat}}$ tends to increase with $M_h$ for $M_h\lesssim10^{13}~M_{\odot}/h$. 

We shall use the trends we have noted in this section in the next few sections, where we build CLF models to populate AGNs in $M_h\gtrsim10^{14},10^{13.5},10^{13},10^{13}~M_{\odot}/h$ haloes at $z\sim 0.6,1.0,1.5,2.0$ respectively.

\section{AGN-Luminosity function}
\label{LMF_sec}
\subsection{Simulations vs. observations}
Filled blue circles and green circles in the top panels of Figure \ref{LMF_sims_vs_obs_fig} show the MBII predictions for the luminosity functions of central and satellite AGNs respectively. We see that the central AGNs are the dominant contribution to the luminosity functions. We can compare the blue circles to the observational measurements from \cite{2009MNRAS.399.1755C} and \cite{2006AJ....131.2766R} shown as open and closed black squares respectively. This shows the range of magnitudes probed by simulations and observations at various redshifts. At $z=0.6,1$, we see that there is a significant (albeit partial) overlap between the simulations and the magnitude range of the observations over $-20 \gtrsim M_g \gtrsim -24$ and there is reasonable agreement between the two. However at $z\sim1.5,2$, the magnitude range of simulated and observed AGNs do not overlap because the number of AGNs in the simulation rapidly declines.

\subsection{CLF modeling of the AGN luminosity function}
Because the simulated and observed luminosity functions do not overlap, we use CLF modeling to extend the range of our predictions to higher luminosities. We consider only central galaxies, thereby assuming that the centrals continue to dominate over satellites at magnitudes brighter than those probed by the simulation ($M_g\lesssim-24$).   

In section \ref{MB2_CLF_sec}, we noted that MBII constrains the dependence of $L^*_{\mathrm{cen}}$ on $M_h$ for halo masses that can be effectively probed by the simulation ($M_h\lesssim10^{14},10^{13.5},10^{13},10^{13}~M_{\odot}/h$ for $z\sim 0.6,1.0,1.5,2.0$ respectively). We find that within the range probed by simulations, $L^*_{\mathrm{cen}}\propto M_h$. In order to reach the observed range of magnitudes, we need to extend the $L^*_{\mathrm{cen}}$ vs. $M_h$ relation to haloes more massive than MBII can effectively probe. We therefore defining a scaling
\begin{equation}
L^*_{\mathrm{cen}} \propto M_h^{\beta_{\mathrm{cen}}}
\end{equation}
for $M_h\gtrsim10^{14},10^{13.5},10^{13},10^{13}~M_{\odot}/h$ for $z\sim 0.6,1.0,1.5,2.0$ respectively. $\beta_{\mathrm{cen}}$ is a power-law exponent which essentially determines the shape of the luminosity function at $M_g \lesssim -24$. 

We determine the values of $\beta_{\mathrm{cen}}$ required to produce a model luminosity function consistent with the observed luminosity functions. The red and green dashed lines (enclosing the shaded blue regions) in the middle panels of Figure \ref{LMF_sims_vs_obs_fig} show the CLF model predictions of the luminosity functions, and we see that they are reasonably consistent with the observed measurements. The values of $\beta_{\mathrm{cen}}$ are listed in the legends; the corresponding $L^*_{\mathrm{cen}}$ vs. $M_h$ relations are shown as red and green dashed lines in the top panels of Figure \ref{Lstar_models_fig}. Note that for $z=0.6$~(the top-left panel), $\beta_{\mathrm{cen}}^{\mathrm{final}} \sim 1$ and is an extrapolation of the best fit line in the simulated regime ($M_h
\lesssim10^{14}~M_{\odot}/h$). This is expected because at $z=0.6$, the simulated and observed luminosity functions partially overlap in their magnitude range and are consistent with each other. At higher redshifts (particularly $z \sim 1.5,2$), $\beta_{\mathrm{cen}}^{\mathrm{final}}$ gradually increases and becomes greater than 1; this makes the $L^*_{\mathrm{cen}}-M_h$ relation somewhat steeper than an extrapolation of the best fit line from the simulated regime. This implies that at $z\sim1.5,2$, a change of slope (compared to simulations) in the halo mass-luminosity scaling (at $M_h\sim10^{13}- 10^{13.5}~M_{\odot}/h$) is required to explain the observed luminosity function at these redshifts. This potentially has implications on the modeling of AGN feedback in MBII. We defer this discussion to a future paper since this is not central to this work but interested readers can refer to Section 8 of \cite{2015MNRAS.450.1349K}. For the purposes of this work, we have now constructed a population of central AGNs with abundances comparable to observations; we can now focus on our main objective, which is to construct a CLF model for satellite AGNs in order to probe the one-halo clustering.

\subsection{Modeling the Satellite AGNs}
\begin{figure*}
\includegraphics[width=\textwidth]{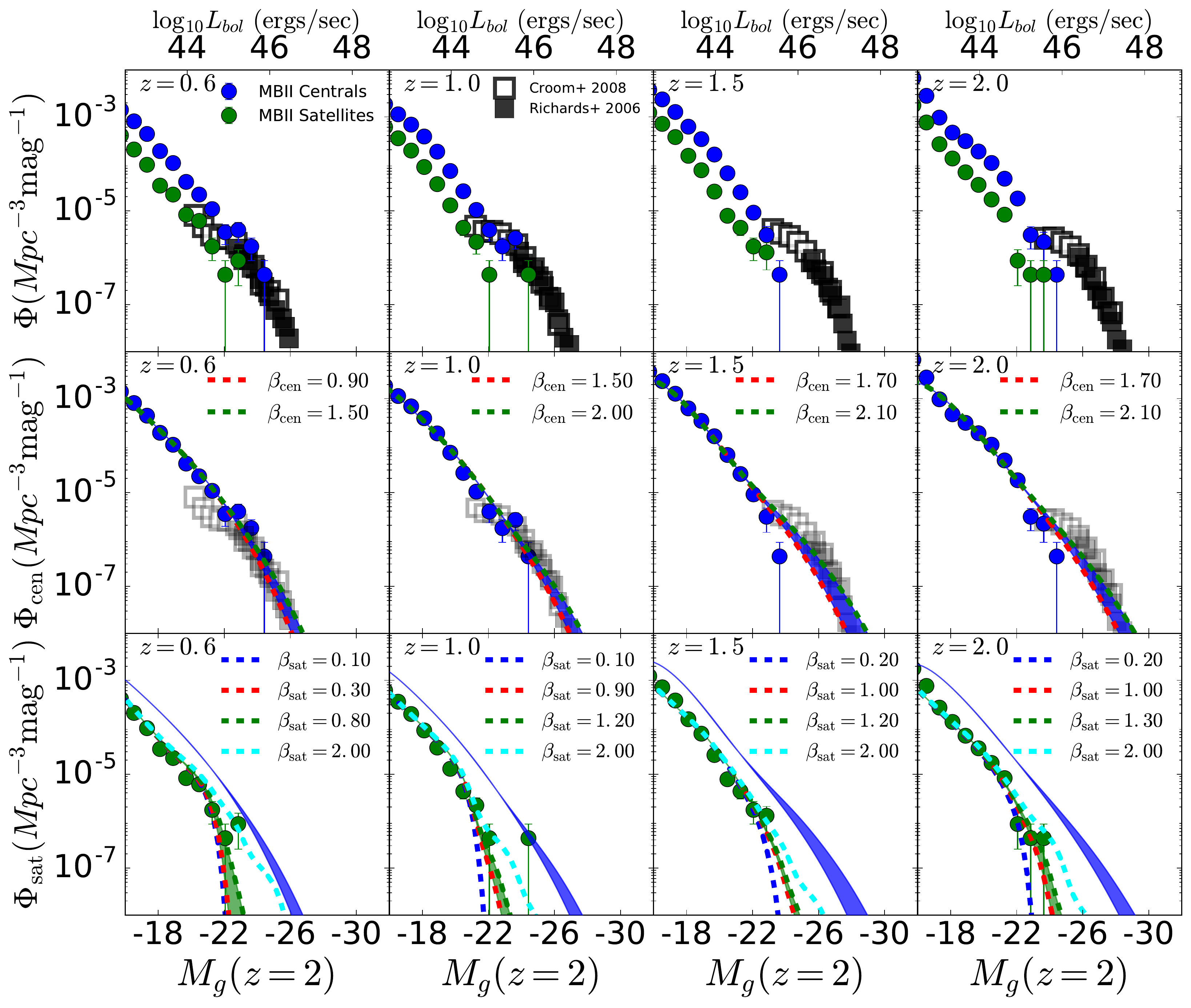}
\caption{\textbf{Top Panels}: Blue and green circles show the luminosity functions (LFs) of central and satellite AGNs respectively; error bars are $1\sigma$ Poisson errors. Filled and open black squares show the observed measurements. \textbf{Middle Panels}: The red and green dashed lines (enclosing the shaded blue region) show the CLF model predictions of central AGN LFs for values of $\beta_{\mathrm{cen}}$ shown in Figure \ref{Lstar_models_fig} (top panels); these models produce LFs consistent with observations. \textbf{Bottom Panels}: Dashed lines show the CLF model predictions of satellite AGN LFs for the values of $\beta_{\mathrm{sat}}$ shown in the bottom panels of Figure \ref{Lstar_models_fig}. The red and green dashed lines (enclosing the shaded green region) correspond to the range of models which produce a small-scale clustering prediction consistent with current constraints (as shown in Figure \ref{average_projected_clustering}).}
\label{LMF_sims_vs_obs_fig}
\end{figure*}

For the CLF modeling of satellite AGNs, we adopt a similar approach to centrals i.e. we define a scaling relation  
\begin{equation}
L^*_{\mathrm{sat}} \propto M_h^{\beta_{\mathrm{sat}}}
\end{equation}
for $M_h\gtrsim10^{14},10^{13.5},10^{13},10^{13}~M_{\odot}/h$ for $z\sim 0.6,1.0,1.5,2.0$ respectively; $\beta_{\mathrm{sat}}$ is a power law exponent which determines the satellite LF at $M_g\lesssim-22$.  However, unlike the centrals, there are no observational measurements of the satellite AGN luminosity functions with which to constrain $\beta_{\mathrm{sat}}$. Therefore, we consider a family of CLF models with various possible values $\beta_{\mathrm{sat}}$, which are shown as dashed lines in the bottom panels of Figure \ref{Lstar_models_fig}. The dashed lines in the lowermost panels of Figure \ref{LMF_sims_vs_obs_fig} show the corresponding values of the satellite LFs; as we can see, we consider models wherein the satellites do not overshoot the central LFs. 

In the next section, we shall look at the one-halo clustering predicted by the CLF models for $g<20.85$ quasars, and compare to the observational measurements of E17.    

\section{Small-scale clustering: Comparison with observational constraints}
\label{one_halo_sec}
\begin{figure*}
\includegraphics[width=\textwidth]{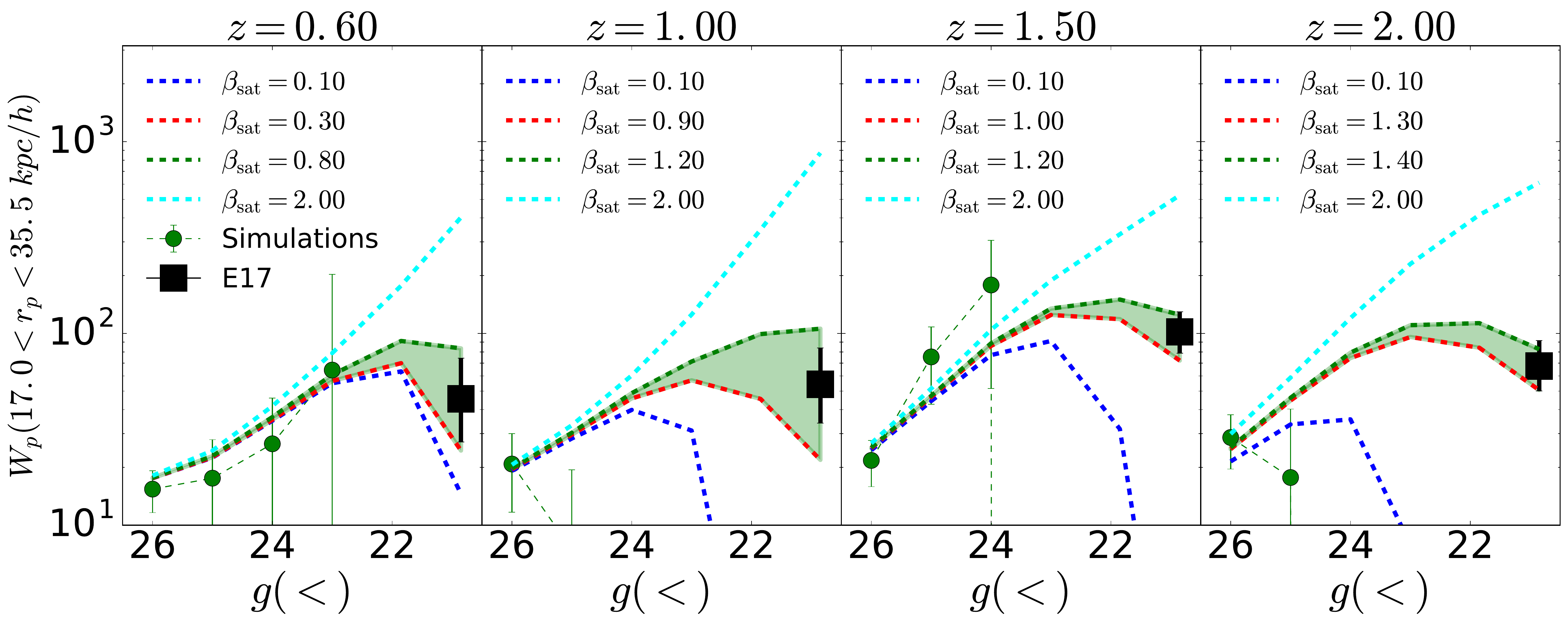}
\caption{$W_p$ is the volume-averaged projected correlation function averaged over $17.0< r_p<36.6\ {\rm kpc}/h$ for quasars brighter than a given magnitude threshold, which we denote by `$g(<)$'. The dashed lines correspond to predictions from the different CLF models (corresponding to the different values of $\beta_{\mathrm{sat}})$ that are considered in the bottom panels of Figures \ref{Lstar_models_fig} and \ref{LMF_sims_vs_obs_fig}.  The black squares correspond to the observational constraints from E17 at $g<20.85$. The red and green dashed lines (enclosing the green shaded region) correspond to the range of models that is consistent with the observed small-scale clustering measurements. The filled circles are predictions from MBII with error bars showing $1\sigma$ Poisson errors.}
\label{average_projected_clustering}
\end{figure*}

E17 measured the volume-averaged projected correlation function ($\bar{W}_p$) over scales of $17.7$--$36.6\ \mathrm{kpc}/h$ for quasar pairs with velocity differences of $<2000$ km/sec. We compute the $\bar{W}_p(17.7$--$36.6\ \mathrm{kpc}/h)$ predicted by our CLF model~(using Section 2.2: Method 2), which we present as dashed lines for various $\beta_{\mathrm{sat}}$ values in Figure \ref{average_projected_clustering}. The $g$ band apparent magnitudes are obtained from $M_g(z=2)$ using Eq.~(4) of \cite{2016A&A...587A..41P}
\begin{equation}
    M_g(z=2)=g-d_m(z)-(K(z)-K(z=2))
\end{equation} 
where $d_m(z)$ is the distance modulus and $K(z)$ is the k-correction adopted from \cite{2013ApJ...768..105M}.
At fainter magnitudes ($23\lesssim g \lesssim 26$), these dashed lines converge and are reasonably consistent with the simulation predictions (shown as filled green circles). For $g<20.85$, we see that the different models can predict a range of clustering amplitudes. The black squares show the measurements of E17 for their sample of quasar pairs with $g<20.85$. We find that in all redshift bins, there is a set of $\beta_{\mathrm{sat}}$ values (shown as red and green dashed lines that enclose the shaded green region) which predict a clustering amplitude consistent with the measurements of E17. 

\section{Implications of current observational constraints}

\label{redshift_evolution_sec}
\begin{figure}
\includegraphics[width=8cm]{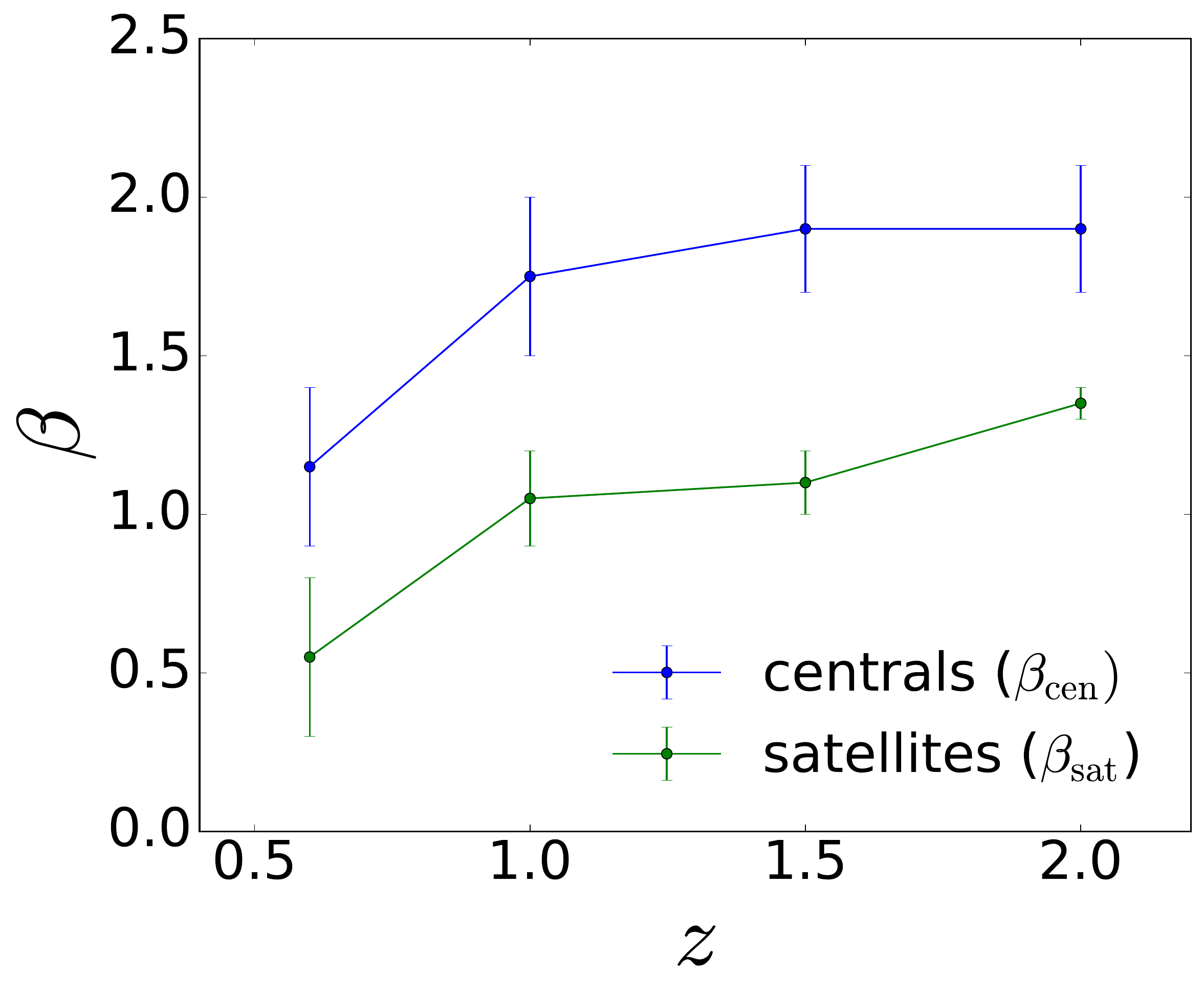}
\caption{Redshift evolution of $\beta_{\mathrm{cen}}$ and $\beta_{\mathrm{sat}}$ from $z=0.6$ to $z=2$. Error bars correspond to the range of values obtained from the current observational constraints on the luminosity functions (for $\beta_{\mathrm{cen}}$) and small-scale clustering (for $\beta_{\mathrm{sat}}$). 
\label{beta}}
\end{figure}
\begin{figure}
\includegraphics[height=7.5cm]{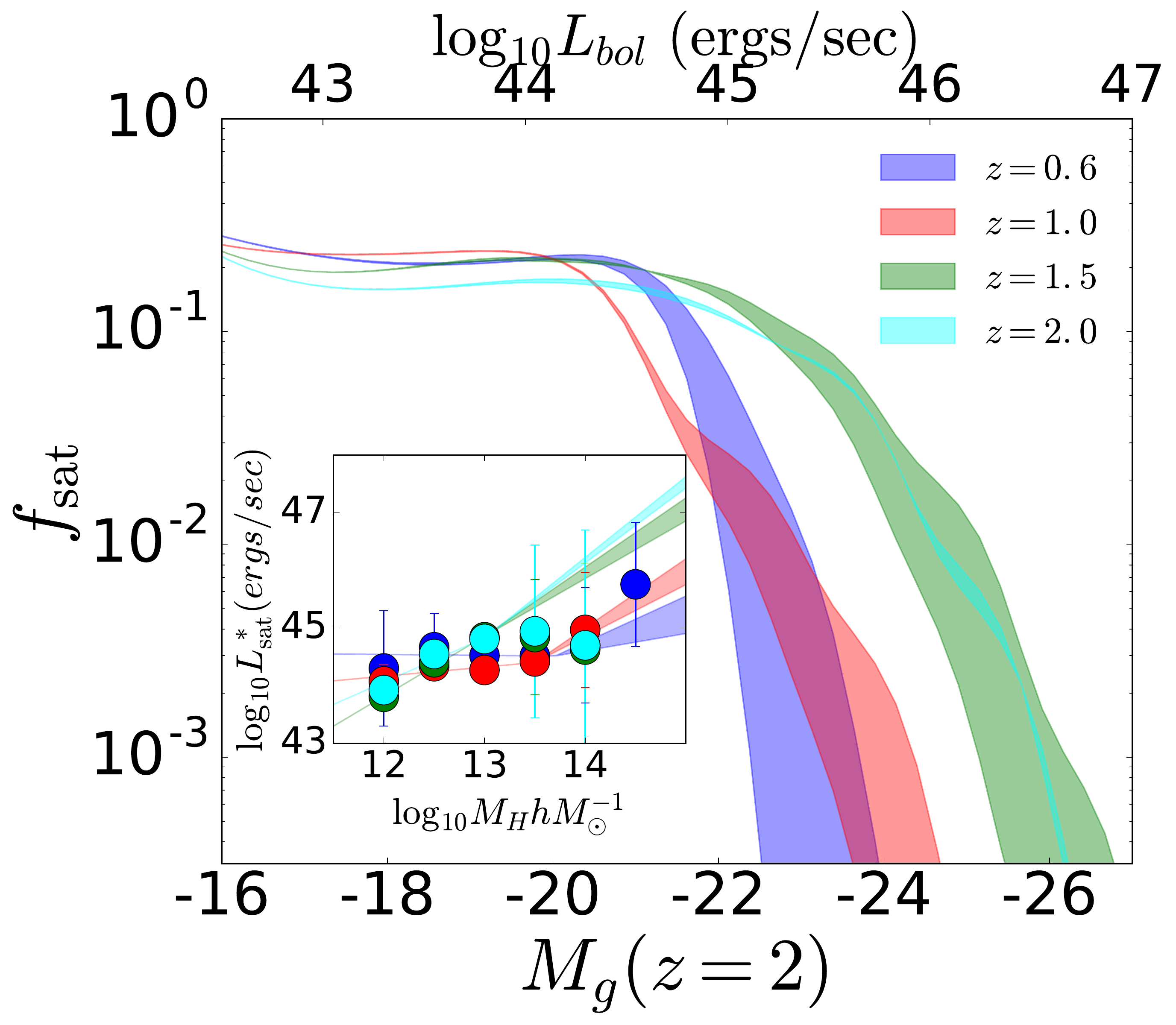}\\
\caption{\textbf{Main Panel}: Shaded regions show the redshift evolution of the satellite fraction from $z\sim0.6$ to $z\sim2$ predicted by our CLF model. \textbf{Inset panel}: Shaded regions show the redshift evolution of the $L^*_{\mathrm{sat}}$ vs. $M_h$ relation from $z\sim0.6$ to $z\sim2$. Filled circles show the simulation predictions.}
\label{z_evo_sat_frac}
\end{figure}
We now select the CLF models which are consistent with observed small-scale clustering constraints and investigate the implications of these models for the AGN population. 

The green circles in Figure \ref{beta} show $\beta_{\mathrm{sat}}$ as a function of redshift for our final CLF model for satellite AGNs. The observed clustering measurements imply $\beta_{\mathrm{sat}}>0$, which means that $L^*_{\mathrm{sat}}$ is positively correlated with halo mass for massive ($M_h\gtrsim10^{14}~M_{\odot}/h$) haloes. A possible physical origin of the positive correlation is the merger of gas-rich massive haloes as triggers of AGN activity (as was seen for the host haloes of the simulated pairs). Higher $M_h$ implies more incoming gas available to fuel the AGNs; as a consequence, the AGNs can reach higher luminosities (higher $L_{\mathrm{sat}}$). In addition, the scaling between $L^*_{\mathrm{sat}}$ and $M_h$ evolves from $L^*_{\mathrm{sat}}\propto M_h^{1/2}$ at $z\sim0.6$ to $L^*_{\mathrm{sat}}\propto M_h$ at $z\sim2$, thereby becoming steeper at higher redshifts for $M_h\gtrsim 10^{14}~M_{\odot}/h$ haloes. This is expected if the halo merger rate increases with redshift, which is indeed the case \citep{2010MNRAS.406.2267F}. Thus, while we model the one-halo clustering as originating primarily from central-satellite quasar pairs, it is likely that these are actually central AGNs that recently became "satellites" after their host haloes underwent a merger. 

We calculate the AGN satellite fraction as defined by $f_{\rm sat}\equiv \Phi_{\mathrm{sat}}/(\Phi_{\mathrm{sat}}+\Phi_{\mathrm{cen}})$. Figure \ref{z_evo_sat_frac} shows the redshift evolution of $f_{\mathrm{sat}}$ from $z\sim0.6$ to $z\sim2$. For $M_g\gtrsim -22$ satellites, the satellite fraction is $\sim 20$--$30\%$ at all redshifts. For $M_g<-22$ satellites, the satellite fraction is $10~\%$ for $z\sim1.5,2.0$ but drops to $0.1~\%$ at $z\sim0.6,1.0$, implying that satellite quasars are significantly more abundant (by factors of $10^2$--$10^3$) at $z\sim2.0$ compared to at $z\sim0.6$. In order to explain the significant increase in satellite quasars at $z\sim1.5,2$, we show the evolution of the $L^*_{\mathrm{sat}}$-$M_h$ relation from $z\sim0.6$ to $z\sim2$ in the inset panel of Figure \ref{z_evo_sat_frac} (the shaded regions are from CLF modeling and the filled circles are for simulations). We see that for $M_h\gtrsim 10^{13}~M_{\odot}/h$ haloes, $L^*_{\mathrm{sat}}$ values are significantly higher (by factors of 10--100) at $z\sim1.5,2$ compared to at $z\sim0.6,1$. This implies that satellite fractions of $M_g\lesssim -22$ quasars increase significantly from $z\sim 0.6$ to $z\sim2$.

Finally, it is worth noting that our CLF models, when compared to the observed measurements, imply halo masses of $\gtrsim10^{14}~M_{\odot}/h$ for the SDSS $g<20.85$ quasar pairs, which is consistent with our simulated quasar pairs. 

\section{Forecasts for EBOSS quasars}
\label{forecasts_sec}
\begin{figure}
\includegraphics[width=8cm]{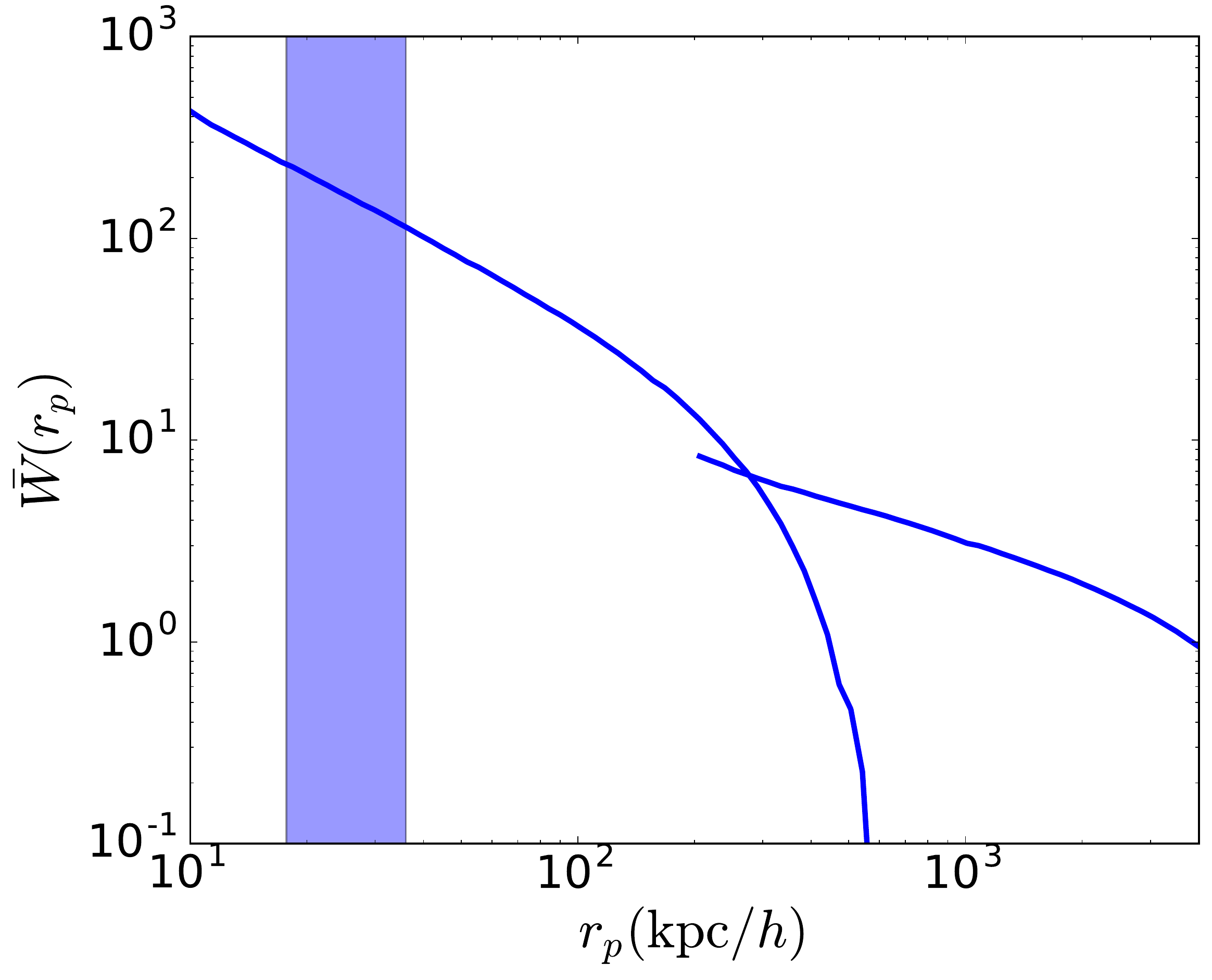}
\caption{Solid lines show the CLF model predictions for the one-halo and two-halo contributions to the AGN clustering for the eBOSS sample ($g<22$ quasars) at a target redshift of $z=1.5$. The shaded region show the scales probed in E17 i.e. $17.7\lesssim r_p \lesssim 36~\mathrm{kpc}/h$.}
\label{eboss}
\end{figure}

We now use our CLF model to make predictions for spatial clustering of quasars in the ongoing eBOSS survey \citep{2016AJ....151...44D}. eBOSS is expected to detect $\gtrsim500,000$ quasars with $g<22$ between $z\sim0.9$ to $2.2$ \citep{2015ApJS..221...27M}. Figure \ref{eboss} presents the one-halo and two-halo contributions of $W_p$ at a target redshift of $z=1.5$. We note that the one-halo term ($r_p\lesssim1~\mathrm{Mpc}/h$) is somewhat enhanced and steeper compared to the two-halo term ($r_p\gtrsim1~\mathrm{Mpc}/h$), which is expected given the high satellite fractions at $z\sim1.5,2$ discussed in the previous section. Strong clustering at small scales, consistent with a high satellite fraction, has been reported in multiple measurements of small-scale quasar clustering \citep[e.g.][]{2006AJ....131....1H,2007ApJ...658...99M, 2008ApJ...678..635M, 2012MNRAS.424.1363K, 2017MNRAS.468...77E}.

We focus on scales targeted by E17 i.e. $17.7\lesssim r_p\lesssim 36~\mathrm{kpc}/h$~(the shaded region in Figure \ref{eboss}). At these scales the clustering amplitude $W_p$ of eBOSS quasar pairs is predicted to be $\sim100$--$200$. We can use the clustering amplitude to calculate the expected number of quasar pairs within the survey area of eBOSS i.e. $\sim7500~\mathrm{deg}^2$ \citep{2016AJ....151...44D}. For a redshift bin-width of $0.46$~(the same as the sample of E17) centered at $z=1.5$, we expect $~200$--$500$ quasar pairs at scales of $17.7\lesssim r_p\lesssim 36~\mathrm{kpc}/h$. This is $\gtrsim10$ times larger than the E17 sample. Additionally, our CLF model predicts that binary quasars in the eBOSS-CORE sample are expected to have host halo masses of $\gtrsim10^{13}~M_{\odot}/h$, which is consistent with the host halo masses of our simulated $g<22$ satellite AGNs.  

\section{Summary and Conclusion}
\label{results_sec}
In this work, we have analysed recent high-precision small-scale clustering measurements of SDSS quasar pairs ($g<20.85$) using the MBII simulation and CLF modeling. Within the MBII volume, there are 3 quasar pairs at these high luminosities at distances of $\sim 1$--$4$ Mpc. Within $M_h\sim 1$--$3\times10^{14}~M_{\odot}/h$ haloes, they reside in separate central and satellite galaxies both of which have stellar masses $M_*\gtrsim1$--$3\times10^{12}~M_{\odot}$ and black halo masses $1$--$3\sim 10^9~M_{\odot}/h$. These galaxies and black holes are among the most massive in the entire simulation. The growth history of these quasars revealed that for all three of the pairs, their progenitor black holes lived in massive $\sim10^{14}~M_{\odot}/h$ haloes which recently merged thereby triggering their AGN activity. This hints at the possibility that rare events such as binary quasars likely originate from mergers of rare massive haloes which have enough gas to sustain these systems.

Next, we look at the small-scale clustering of binary quasars. Given that these quasar pairs are so rare, we use CLF modeling built using the MBII simulation to analyse their one-halo clustering. For the MBII AGNs, we find that CLFs of centrals and satellites can be well described by log-normal and Schechter distributions respectively. Assuming that the central AGNs dominate over satellites in the luminosity function, we built a CLF model for central AGNs which predicts an AGN luminosity function consistent with that of the simulated MBII AGNs as well as observed quasars. For the satellite AGNs, based on the trends exhibited by the CLFs of MBII AGNs in $M_h\lesssim10^{14}~M_{\odot}/h$ haloes, we considered a family of CLF models to extrapolate to $M_h\gtrsim10^{14}~M_{\odot}/h$ haloes and make predictions for the small-scale (one-halo) clustering; in particular, the central-satellite term. We compare these predictions to small-scale clustering measurements from observations at $\sim 25~\mathrm{kpc}$ scales, and arrive at a final model which is consistent with the measurements. Our key parameter of interest is $L^*_{\mathrm{sat}}$, which is the maximum luminosity of a satellite AGN in a halo of a given mass $M_h$. Constraining our model with the observed measurements leads to three interesting findings about the $L^*_{\mathrm{sat}}-M_h$ relation:

\begin{enumerate}
 \item $L^*_{\mathrm{sat}}$ has significant positive correlation with $M_h$ for $M_h \gtrsim 10^{14}~M_{\odot}/h$ haloes.
 
 \item The correlation gets stronger with redshift evolving from $L^*_{\mathrm{sat}}\sim M_h^{1/2}$ at $z=0.6$ to $L^*_{\mathrm{sat}}\sim M_h$ at $z=2$. 
 
 \item For fixed halo mass, $L^*_{\mathrm{sat}}$ steeply increases (by 2--3 orders of magnitude) from $z\sim0.6$ to $z\sim2$ for $M_h\gtrsim 10^{14}~M_{\odot}/h$ haloes. This leads to a significant increase in the AGN satellite fraction from $f_{\mathrm{sat}} \sim 10^{-3}$ at $z=0.6$ to $f_{\mathrm{sat}} \sim 10^{-1}$ at $z=2$ for $M_g \lesssim -22$ quasars.
\end{enumerate}

These findings are consistent with a scenario where binary quasars are triggered by mergers of massive $\gtrsim 10^{14}~M_{\odot}/h$ haloes (as seen for our simulated pairs). In this scenario, a merger of two such haloes can funnel a significant amount of gas to a black hole in a relatively short time, thereby increasing the activity in satellite AGNs to $\sim 10^{44.5-45}~\mathrm{ergs/sec}$, making them $\sim 10^2$--$10^3$ times more luminous compared to a typical satellite AGN (with mean luminosity of $\sim 10^{41.5-42}~\mathrm{ergs/sec}$, see Figure \ref{scaling_relations_fig}). Therefore, these mergers potentially affect the most luminous ``edge'' of the satellite CLF, quantified by $L^*_{\mathrm{sat}}$. To explain point i) in our summary, above, mergers of increasingly massive haloes are accompanied by increasing amounts of incoming gas now available to feed the black holes. Thus, quasars can reach increasingly higher luminosities with increasing halo mass, thereby leading to a positive correlation between $L^*_{\mathrm{sat}}$ and $M_h$. Points ii) and iii) can then be explained by the fact that the merger rate increases with redshift \citep{2010MNRAS.406.2267F}. Higher numbers of satellites at $z\sim1.5,2$ leads to enhanced one-halo clustering (at $r_p\lesssim 1~\mathrm{Mpc}/h$).

Finally, for the ongoing eBOSS-CORE sample ($g<22$), we predict a small-scale clustering amplitude of $\bar{W}_p \sim 100$--$200$ at $17.7\lesssim r_p \lesssim 36~\mathrm{kpc}/h$. This  corresponds to $\sim 200$--$500$ pairs (with separations of $< 2000\,{\rm km/s}$) expected at these scales at $z\sim1.5$. Furthermore, these pairs are expected to live in haloes with $M_h\gtrsim10^{13}~M_{\odot}/h$ and have black-hole masses $M_{bh}\gtrsim10^{8}~M_{\odot}/h$. This predicted sample is $\gtrsim10$ times larger than the size of current samples of quasar pairs.

\section{General remarks and Future work}
\label{remarks_sec}
Our work demonstrates that hydrodynamic simulations are invaluable tools to study properties of AGN and quasar populations. In particular, simulations such as MBII can probe faint AGNs ($M_g\gtrsim -18$) which cannot be accessed by observations because their AGN activity is masked by the luminosity of star formation activity within the host galaxy. CLF modeling was a crucial tool in our work to establish the link between faint ($M_g\gtrsim -18$) AGNs (which are difficult to access by observations) to bright ($M_g\lesssim -22$) quasars (which are difficult to access by simulations). It helped us build a model for the AGN-halo connection across a very wide range of luminosities ($-16\gtrsim M_g\gtrsim -32$) and study the redshift evolution of quasars from $z=0.6$ to $z=2$.

It is however important to recognize that our CLF modeling does rely on some implicit assumptions about the AGN population. They are as follows:
\begin{itemize}
    \item We assumed that central AGNs are the dominant contributor to the luminosity function across the entire range of AGN luminosities ($-16\gtrsim M_g\gtrsim -32$). This enabled us to use the observed quasar luminosity function to constrain the CLF model for central AGNs in haloes not well probed by MBII simulation i.e. $M_h\gtrsim 10^{14} M_{\odot}/h$.
    
    \item We assumed that the normalization~($f_{\mathrm{active}}$) of the central CLFs continues to be 1 for haloes with masses larger than $\sim10^{14.5}~M_{\odot}/h$ which are too rare to be probed by MBII.   
    
    \item We assumed that the model CLFs for satellite and central AGNs follow Schechter and log-normal distributions respectively in $M_h\gtrsim 10^{14} M_{\odot}/h$ haloes (which are not well probed by simulations).
    
    \item Among the CLF model parameters, $L^*_{\mathrm{cen}}$ and $L^*_{\mathrm{sat}}$ (and their dependence on halo mass at $M_h\gtrsim 10^{14} M_{\odot}/h$) primarily determine the quasar luminosity function at $M_g<-22$ and the small-scale clustering of $g<20.85$ quasars; therefore we only allowed 2 parameters ($\beta_{\mathrm{cen}}$ and $\beta_{\mathrm{sat}}$) to vary in our initial family of CLF models. The modeling of other parameters ($\sigma_{\mathrm{cen}}$, $\alpha_{\mathrm{sat}}$, $Q_{\mathrm{sat}}$) were fixed based on the trends seen in the MBII simulation (see Appendix \ref{modeling_CLFs} for details).
\end{itemize}

Relaxing one or more of the above assumptions greatly increases the complexity of the problem because the parameter space becomes large (6 parameters). Future work should involve the use of more sophisticated techniques (Markov chain Monte Carlo for example) to constrain the CLF models without depending on one or more of these assumptions, and therefore look for potential degeneracies in the modeling.

Efforts will also continue to exploit the rapid progress in computational power to push hydrodynamic simulations to larger volumes. This may ultimately allow simulations to directly probe the statistical properties of close pairs of quasars. 

\section*{Acknowledgments}
AKB, TD, SE and ADM were supported by the National Science Foundation through grant number 1616168. TDM acknowledges funding from NSF
ACI-1614853, NSF AST-1517593, NASA ATP NNX17AK56G and NASA ATP 17-0123. The \texttt{BLUETIDES} simulation was run on the BlueWaters facility at the National Center for Supercomputing Applications. ADM also acknowledges support through the U.S. Department of Energy, Office of Science, Office of High Energy Physics, under Award Number DE-SC0019022. 
AKB is thankful to Duncan Campbell and Francois Lanusse for useful discussions.

\appendix

\section{Black hole duty cycles in MBII}
\label{black_hole_duty_cycles}
\begin{figure*}
\addtolength{\tabcolsep}{-8pt} 
\begin{tabular}{cc}
\includegraphics[width=8cm]{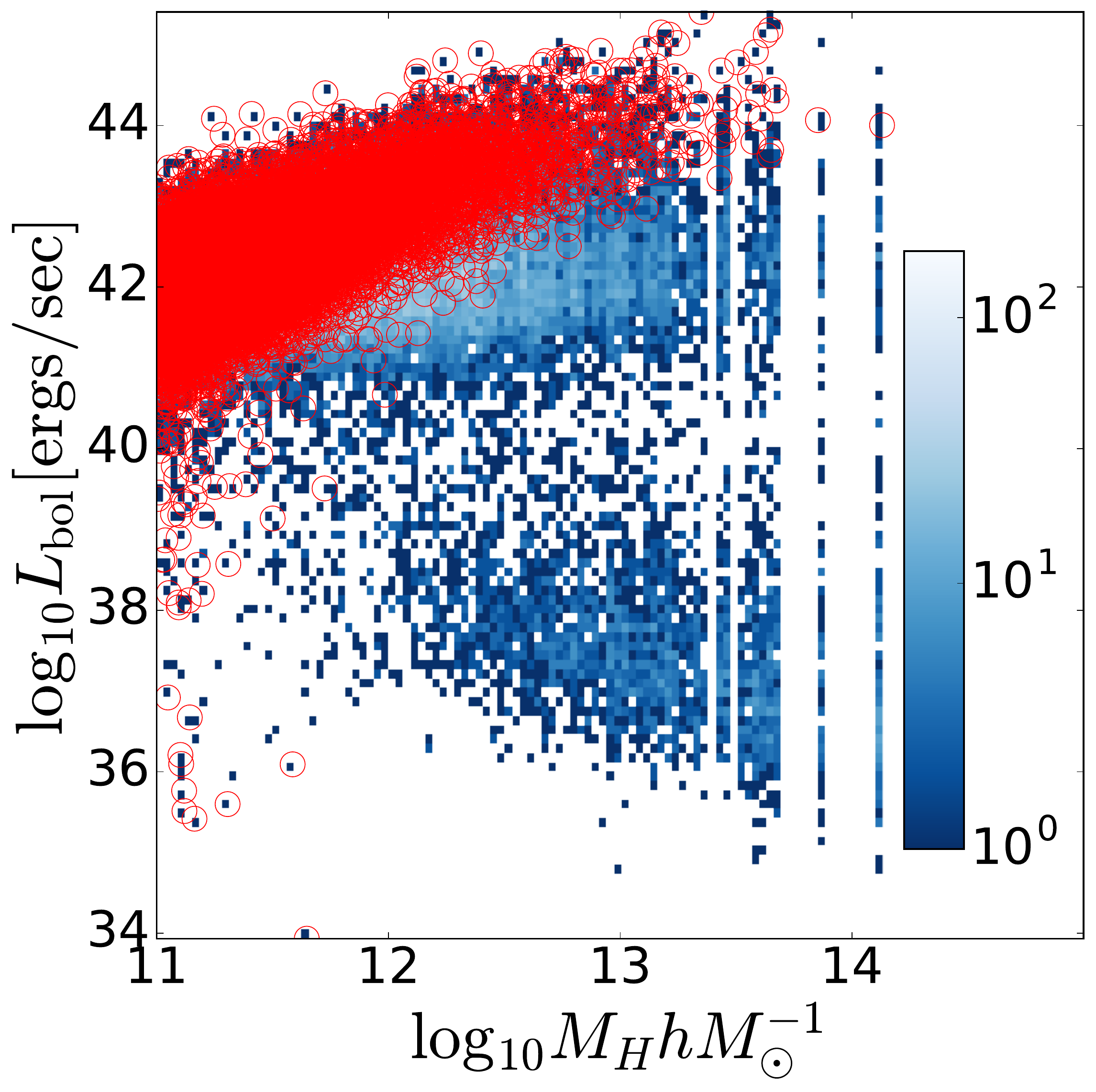}&
\hspace{0.5cm}\includegraphics[width=8.5cm]{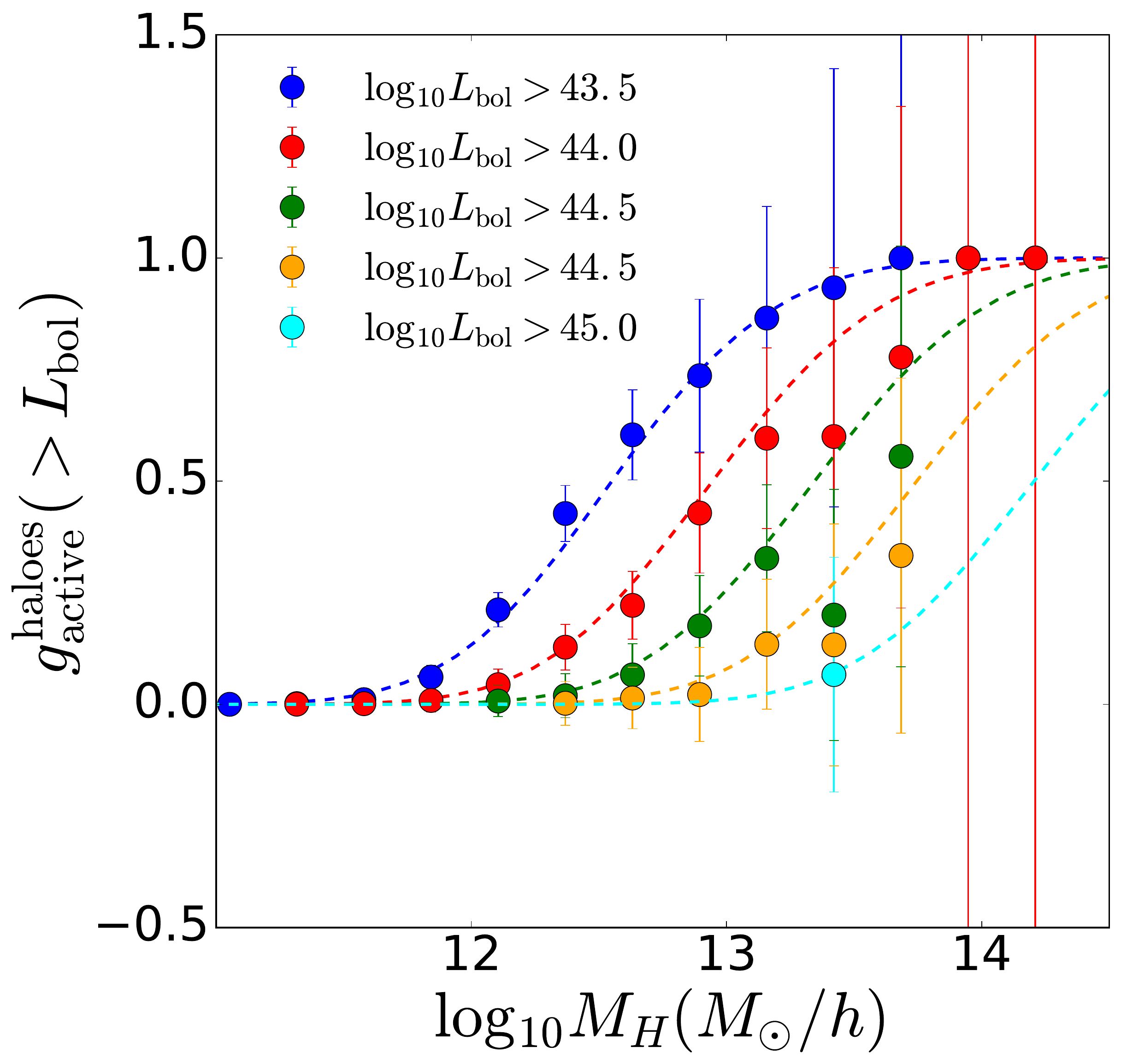}
\end{tabular}
\caption{\textbf{Left Panel}: Blue histograms show the bolometric luminosity vs. halo mass relation of the total black hole population at $z=1.5$ snapshot of the MBII simulation. The red circles indicate \textit{central} black hole. 
\textbf{Right Panel}: $g^{\mathrm{haloes}}_{\mathrm{active}}$ is the fraction of haloes hosting an \textit{active} \textit{central} AGN above various possible bolometric luminosity thresholds~(see legend). The filled circles correspond to MBII predictions at $z=1.5$ snapshot with error bars showing Poisson errors. The dashed lines show predictions for our CLF model.}
\label{duty_cycle}
\end{figure*}

AGN activity depends on the density/ angular momentum of the surrounding gas; AGNs are therefore not active at all times. As a result, not all haloes may necessarily host \textit{active}~(as defined in Section \ref{central_satellite_identification}) black holes~(independent of halo mass). In such a case, the overall normalization $f_{\mathrm{active}}$ of the central AGN CLF can assume any value between 0 to 1. To investigate this further, we plot in Figure \ref{duty_cycle}: left panel the scaling relations~(blue histograms) between bolometric luminosity and host halo mass for the \textit{complete} population of MBII black holes at $z=1.5$. The \textit{central} black holes are shown as red circles. As expected, there are a number of \textit{inactive} black holes which we do not consider in our analysis. However, if we look at the \textit{central} black hole population~(red circles), all of them are \textit{active} for \textit{sufficiently massive}~($M_h\gtrsim10^{12}~M_{\odot}/h$) haloes. In other words, while \textit{inactive} black holes are certainly present, for all \textit{sufficiently massive}~($M_h\gtrsim10^{12}~M_{\odot}/h$) haloes there is always \textit{at least one} black hole~(\textit{central}) inside each halo which is \textit{active}.  

Figure \ref{duty_cycle}: right panel shows the fraction~($g^{\mathrm{haloes}}_{\mathrm{active}}$) of haloes hosting an \textit{active} central black hole as a function of halo mass~$M_h$. As expected from Figure \ref{duty_cycle}: left panel, we find that for all luminosity thresholds, $g^{\mathrm{haloes}}_{\mathrm{active}}$ goes to 1 for large enough halo masses~(dashed lines show the predictions of $g^{\mathrm{haloes}}_{\mathrm{active}}$ from our CLF model). This implies that the overall normalization for the Conditional Luminosity Function~(CLF) of \textit{active} \textit{central} AGNs is $f_{\mathrm{active}}\sim1$. These findings are also consistent with other hydrodynamical simulations \citep{2012MNRAS.419.2657C,2017MNRAS.466.3331D}. Observational constraints on $f_{\mathrm{active}}$ are not yet firmly established. While there are works \citep{2012ApJ...755...30R,2018MNRAS.477...45M} which   explain observed AGN clustering with $f_{\mathrm{active}}=1$, and there are also works \citep{2015MNRAS.446.1874L} which explain gravitational lensing measurements from X-ray AGNs with $f_{\mathrm{active}}<1$. This is likely due to potential degeneracies between HOD parameters as discussed in Section \ref{remarks_sec}, which will likely be resolved with stronger constraints on clustering/ lensing statistics from ongoing and future surveys such as eBOSS and DESI.

\section{modeling CLFs} 
\label{modeling_CLFs}
\begin{figure*}
\addtolength{\tabcolsep}{-8pt} 
\begin{tabular}{ccc}
\includegraphics[width=6cm]{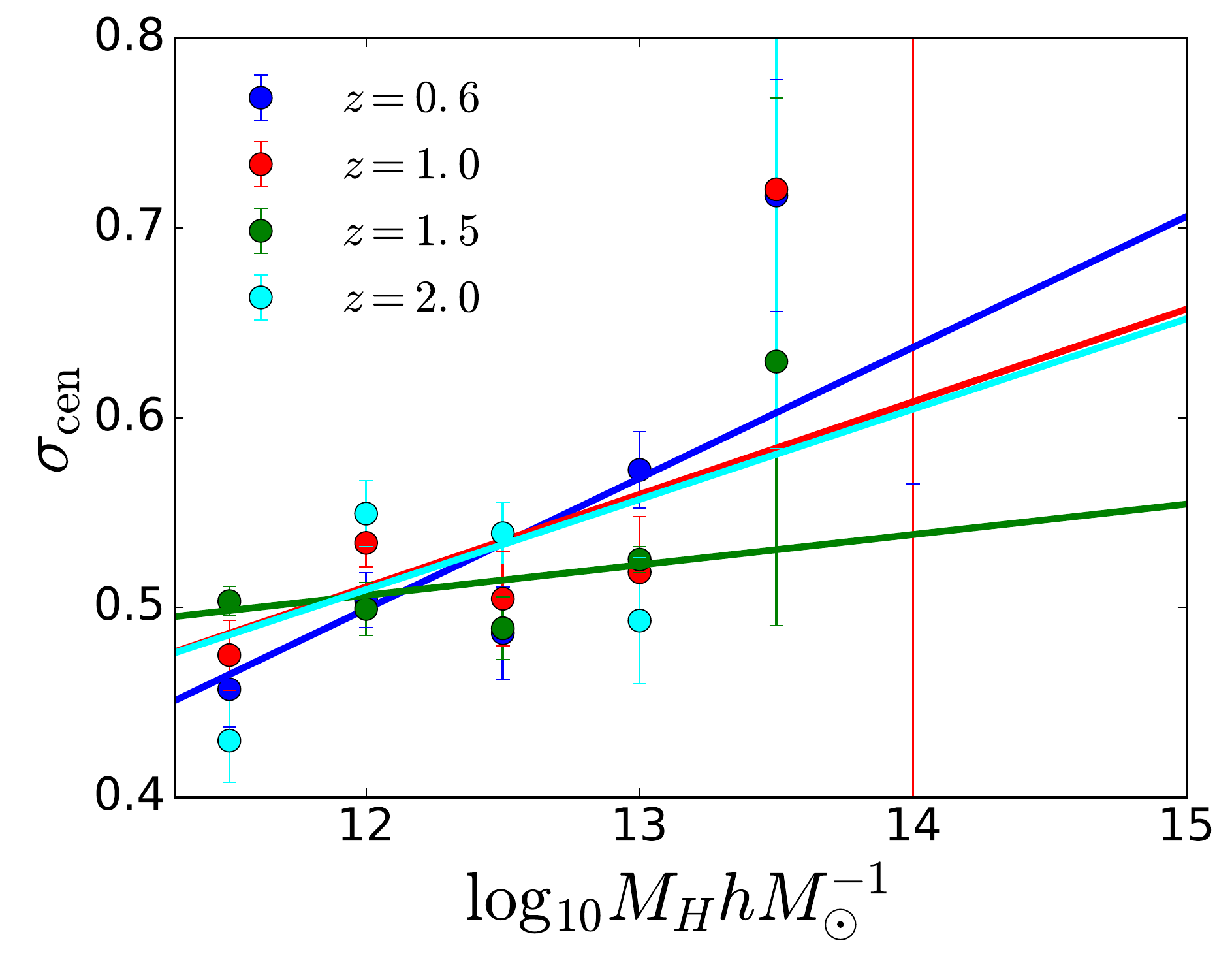}&
\includegraphics[width=6cm]{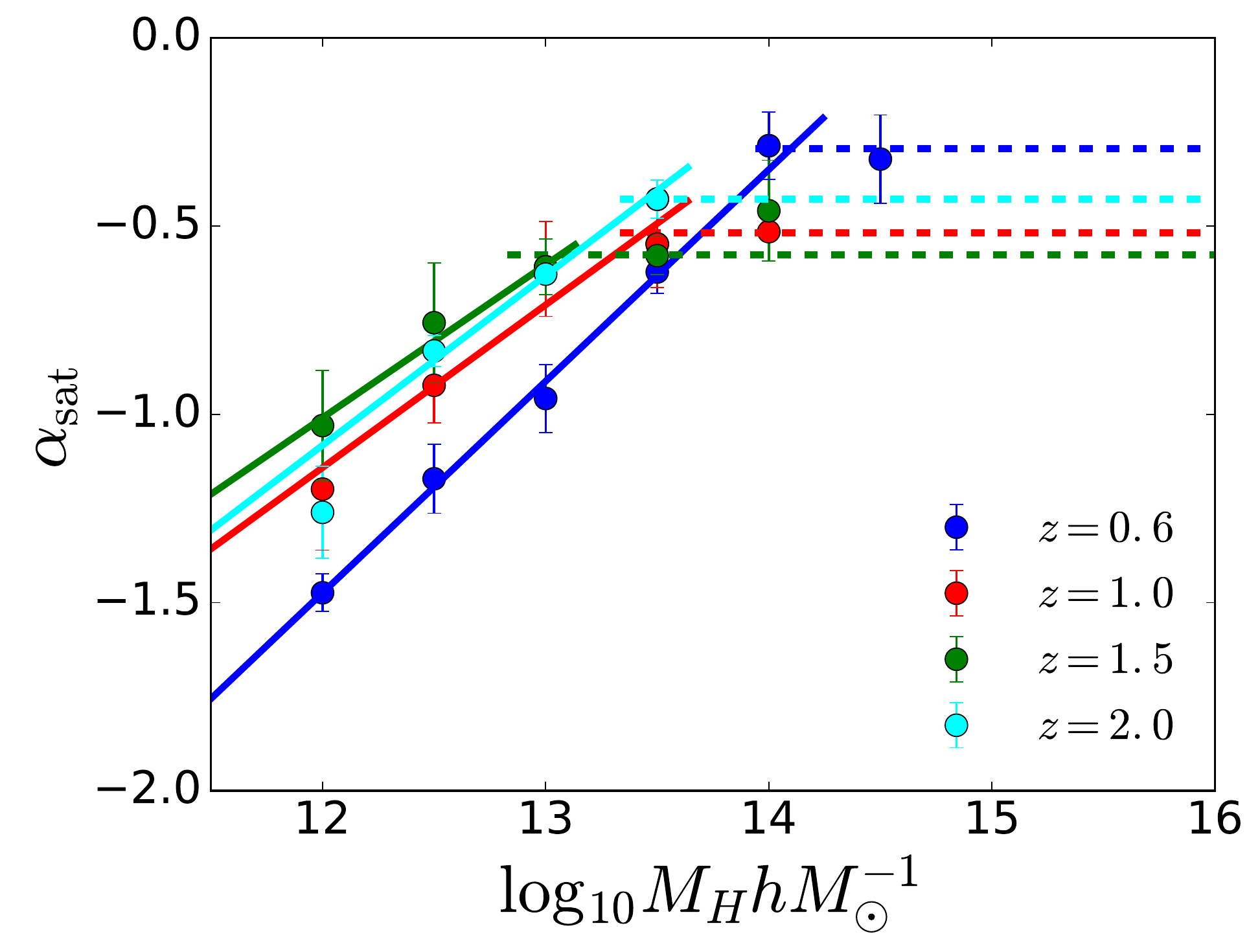}&
\includegraphics[width=6cm]{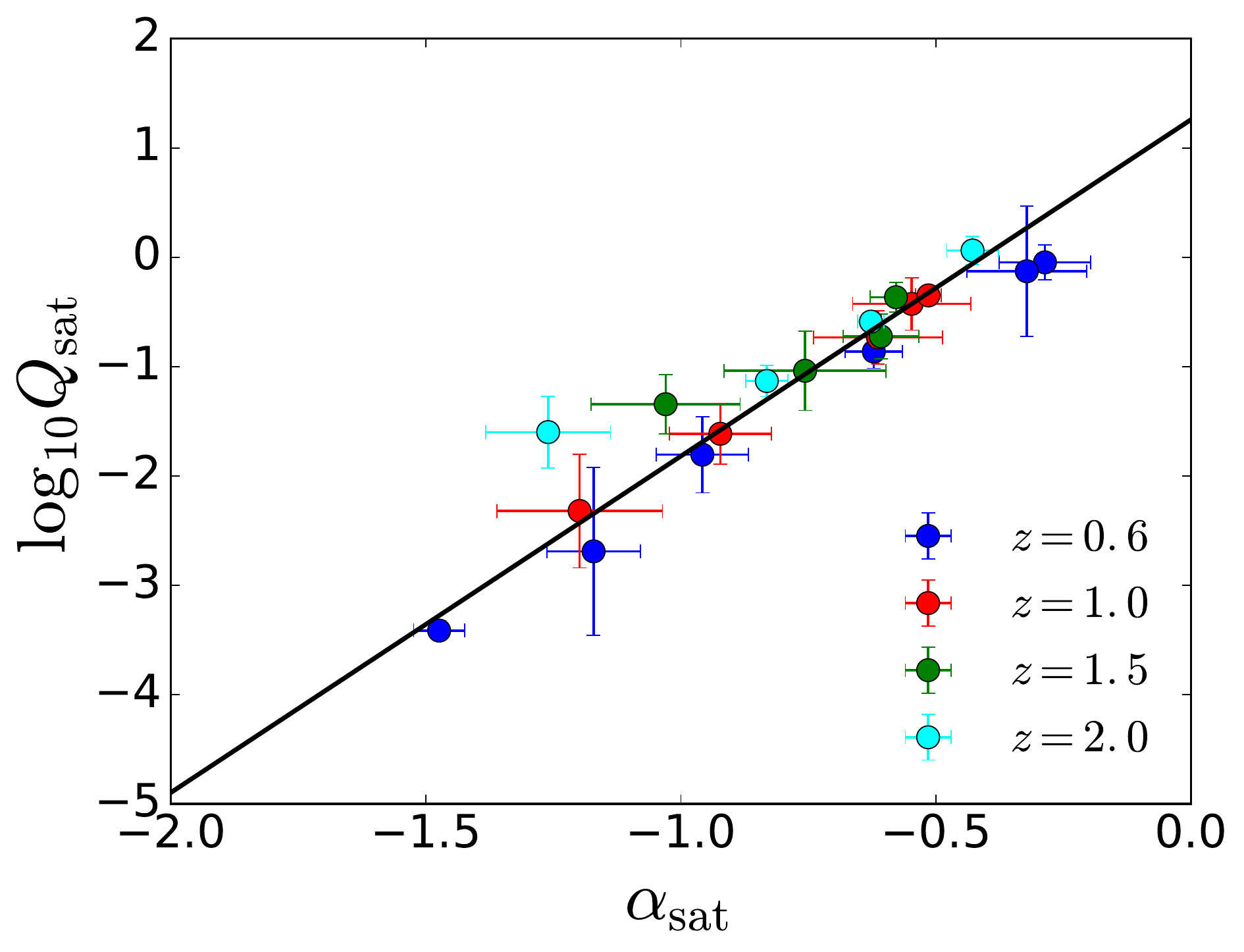}
\end{tabular}
\caption{\textbf{Left Panel}: Filled circles show the best fit values of $\sigma_{\mathrm{cen}}$ for the Schechter fits to CLMFs of satellite AGNs at various redshifts. The error bars show the $1\sigma$ errors. The solid lines show the best fit linear regression over the appropriate mass range.
\textbf{Middle Panel}: Filled circles show the best fit values of $\alpha_{\mathrm{sat}}$ for the Schechter fits to CLMFs of satellite AGNs at various redshifts. The error bars show the $1\sigma$ errors. The solid lines show the best fit linear regression over the appropriate mass range. Dashed lines show the value of $\alpha_{\mathrm{sat}}$ assumed over the corresponding mass range as shown. \textbf{Right Panel}: Filled circles show the correlation between the best fit values of  $Q_{\mathrm{sat}}$ and $\alpha_{\mathrm{sat}}$ at various redshifts. The black solid line shows the best fit linear regression.}
\label{sigma_logN_fig}
\end{figure*}

\begin{table*}
    \centering
    \begin{tabular}{c|c|c|c|c}
    Redshift&$M'_H~(M_{\odot}/h)$&Halo Mass Range&  CLF model for Central AGN & \\
    \hline
        \multirow{2}{*}{0.6}&\multirow{2}{*}{$10^{13.8}$}&$M_h \lesssim M'_h$& $L^*_{cen}= 10^{30.0\pm0.2}M_h^{1.02\pm0.02}$; $\sigma_{cen}= (0.07\pm0.01)\log_{10}{M_h}+(-0.32\pm0.18)$\\
        \addlinespace[0.1cm]
     & &$M_h \gtrsim M'_H $& $L^*_{cen}= 10^{44.1}(M_h-M'_H)^{1.2\pm0.2}$; $\sigma_{cen}= (0.07\pm0.01)\log_{10}{M_h}+(-0.32\pm0.18)$\\
    \hline
        \multirow{2}{*}{1.0}&\multirow{2}{*}{$10^{13.5}$}&$M_h \lesssim M'_h$& $L^*_{cen}= 10^{29.5\pm0.7}M_h^{1.08\pm0.05}$; $\sigma_{cen}= (0.05\pm0.02)\log_{10}{M_h}+(-0.07\pm0.23)$\\
        \addlinespace[0.1cm]
     & &$M_h \gtrsim M'_h$& $L^*_{cen}= 10^{44.1}(M_h-M'_H)^{1.7\pm0.2}$; $\sigma_{cen}= (0.05\pm0.02)\log_{10}{M_h}+(-0.07\pm0.23)$\\
    \hline
        \multirow{2}{*}{1.5}&\multirow{2}{*}{$10^{13}$}&$M_h \lesssim M'_h$& $L^*_{cen}= 10^{30.5\pm0.4}M_h^{1.03\pm0.03}$; $\sigma_{cen}= (0.02\pm0.01)\log_{10}{M_h}+(0.31\pm0.06)$\\
        \addlinespace[0.1cm]
     & &$M_h \gtrsim M'_h$& $L^*_{cen}=10^{43.8}(M_h-M'_H)^{2.0\pm0.4}$; $\sigma_{cen}= (0.02\pm0.01)\log_{10}{M_h}+(0.31\pm0.06)$\\
    \hline
        \multirow{2}{*}{2.0}&\multirow{2}{*}{$10^{13}$}&$M_h \lesssim M'_h$& $L^*_{cen}=10^{26.7\pm1.1} M_h^{1.29\pm0.09}$; $\sigma_{cen}= (0.05\pm0.04)\log_{10}{M_h}+(-0.06\pm0.50)$\\
        \addlinespace[0.1cm]
     & &$M_h \gtrsim M'_h$& $L^*_{cen}=10^{44.2}(M_h-M'_H)^{2.0\pm0.5}$; $\sigma_{cen}= (0.05\pm0.04)\log_{10}{M_h}+(-0.06\pm0.50)$\\
    \hline
    \end{tabular}

\caption{Summary of the CLF model for central AGNs. $L^*_{\mathrm{cen}}$ and $M_h$ are written in units of $\mathrm{ergs/sec}$ and $M_{\odot}/h$ respectively. Error bars are covariance error estimates obtained from the linear fits. }        

\label{summary_table1}
\end{table*}

\begin{table*}
    \centering
    \begin{tabular}{c|c|c|c|c}
    Redshift&$M'_H~(M_{\odot}/h)$&Halo Mass Range& CLF model for Satellite AGN \\
    \hline
        \multirow{2}{*}{0.6}&\multirow{2}{*}{$10^{13.8}$}&$M_h \lesssim M'_h$ & $L^*_{\mathrm{sat}}=10^{45\pm1} M_h^{0.01\pm0.08}$; $\alpha_{\mathrm{sat}}= (0.56\pm0.02)\log_{10}{M_h}+(-8.2\pm0.2)$\\
        \addlinespace[0.1cm]
     & &$M_h \gtrsim M'_h$& $L^*_{\mathrm{sat}}=10^{44.5}(M_h-M_h^')^{0.5\pm0.2}$; $\alpha_{\mathrm{sat}}=-0.29\pm0.01$
     \\
    \hline
        \multirow{2}{*}{1.0}&\multirow{2}{*}{$10^{13.5}$}&$M_h \lesssim M'_h$& $L^*_{\mathrm{sat}}= 10^{42\pm1}~M_h^{0.16\pm0.09}$;  $\alpha_{\mathrm{sat}}= (0.43\pm0.07)\log_{10}{M_h}+(-6.3\pm1.0)$\\
        \addlinespace[0.1cm]
     & &$M_h \gtrsim M'_h$ & $L^*_{\mathrm{sat}}=10^{44.3}(M_h-M_h^')^{1.0\pm0.2}$; $\alpha_{\mathrm{sat}}=-0.51\pm0.01$
     \\
    \hline
        \multirow{2}{*}{1.5}&\multirow{2}{*}{$10^{13}$}&$M_h \lesssim M'_h$& $L^*_{\mathrm{sat}}=10^{36\pm2}M_h^{0.68\pm0.2}$; $\alpha_{\mathrm{sat}}= (0.40\pm0.05)\log_{10}{M_h}+(-5.8\pm0.7)$\\
        \addlinespace[0.1cm]
     & &$M_h \gtrsim M'_h$& $L^*_{\mathrm{sat}}=10^{44.8}(M_h-M_h^')^{1.0\pm0.2}$; $\alpha_{\mathrm{sat}}=-0.57\pm0.02$
     \\
    \hline
        \multirow{2}{*}{2.0}&\multirow{2}{*}{$10^{13}$}&$M_h \lesssim M'_h$& $L^*_{\mathrm{sat}}=10^{37\pm2}M_h^{0.6\pm0.2}$; $\alpha_{\mathrm{sat}}= (0.45\pm0.06)\log_{10}{M_h}+(-6.4\pm0.8)$\\
        \addlinespace[0.1cm]
     & &$M_h \gtrsim M'_h$&$L^*_{\mathrm{sat}}= 10^{44.8}(M_h-M_h^')^{1.2\pm0.1}$; $\alpha_{\mathrm{sat}}=-0.42\pm0.01$
     \\
    \hline
    \end{tabular}

\caption{Summary of the CLF model for satellite AGNs. $L^*_{\mathrm{sat}}$ and $M_h$ are written in units of $\mathrm{ergs/sec}$ and $M_{\odot}/h$ respectively. Error bars are covariance error estimates obtained from the linear fits.}        

\label{summary_table2}
\end{table*}

\subsection{Central AGNs:}
We modeled the central CLF as a log-normal distribution shown in Eq.~(\ref{central_CLF_eqn}) with mean luminosity $L^*_{\mathrm{cen}} (M_h)$ for a given halo mass $M_h$. Section \ref{CLF_sec} discusses the modeling of $L^*_{\mathrm{cen}} (M_h)$. Here, we complete the discussion of CLF modeling of central AGNs by presenting the other parameter $\sigma_{\mathrm{cen}} (M_h)$, which quantifies the scatter of log-normal distribution.   

Filled circles in Figure \ref{sigma_logN_fig}~(left panel) show the best fit values of $\sigma_{\mathrm{cen}}$ in various halo mass bins. We find that the values of  $\sigma_{\mathrm{cen}}$ range from $0.4$--$0.7$ with a mild increase with halo mass. We find no obvious trend in the redshift evolution. We model the mass dependence using a linear regression between $\sigma_{\mathrm{cen}}$ and $\log_{10}{M_h}$, which are shown as solid lines in the left-hand panel of Figure \ref{sigma_logN_fig}.

\subsection{Satellite AGNs}
We modeled the satellite CLF as a Schechter distribution shown in Eq.~(\ref{satellite_CLF_eqn}) with a maximum luminosity $L^*_{\mathrm{sat}} (M_h)$ for a given halo mass $M_h$. Section \ref{CLF_sec} discusses the modeling of $L^*_{\mathrm{sat}} (M_h)$. Here, we complete the discussion of CLF modeling of satellite AGNs by presenting the remaining parameters $\alpha_{\mathrm{sat}} (M_h)$ and $Q_{\mathrm{sat}} (M_h)$. 

$\alpha_{\mathrm{sat}}$ measures the slope of the Schechter distribution for $L_{\mathrm{sat}}<L^*_{\mathrm{sat}}$. Filled circles in Figure \ref{sigma_logN_fig}~(middle panel) show the best fit values of $\alpha_{\mathrm{sat}}$ in various halo mass bins. We find that $\alpha_{\mathrm{sat}}$ increases with $M_h$ as a power law up to $M_h\sim 10^{14},10^{13.5},10^{13},10^{13.5}~M_{\odot}/h$ for $z=0.6,1.0,1.5,2.0$ respectively; the relations are therefore modelled as solid lines as shown in Figure \ref{sigma_logN_fig}~(middle panel). For more massive haloes, simulations indicate that the $\alpha_{\mathrm{sat}}$ vs. $M_h$ relation flattens; in this regime, we simply model $\alpha_{\mathrm{sat}}$ as a constant value independent of $M_h$ shown as dashed lines. This makes sense because otherwise, a naive extrapolation of the solid lines would imply $\alpha_{\mathrm{sat}}>0$ for large enough halo masses; this would invert the slope of the Schechter distribution implying that AGN abundance increases with luminosity, which is somewhat unphysical.

The final parameter to be modelled is $Q_{\mathrm{sat}} (M_h)$, which simply determines the overall normalization of the Schechter function, given the shape parameters $\alpha_{\mathrm{sat}}$ and $L^*_{\mathrm{sat}}$. Interestingly, we find that $Q_{\mathrm{sat}}$ has a tight correlation with $\alpha_{\mathrm{sat}}$ as shown in the right-hand panel of Figure \ref{sigma_logN_fig}. Furthermore, their relation does not exhibit a significant redshift dependence. We therefore simply model the dependence of $Q_{\mathrm{sat}}$ and $\alpha_{\mathrm{sat}}$ as a linear regression shown as the solid black line; the relation is given as:
\begin{equation}
    \log_{10}{Q_{\mathrm{sat}}}=(3.0 \pm 0.1)\alpha_{\mathrm{sat}}+(1.25 \pm 0.1)
    \label{for_Qsat}
\end{equation}
where the uncertainties in the coefficients are $1\sigma$ errors in the regression. Eq.~(\ref{for_Qsat}) along with Tables A1 and A1 summarize our final CLF model parameters.

\section{Radial profiles of satellite AGNs in MBII}
\label{satellite_profile_sec}
\begin{figure}
\includegraphics[width=8cm]{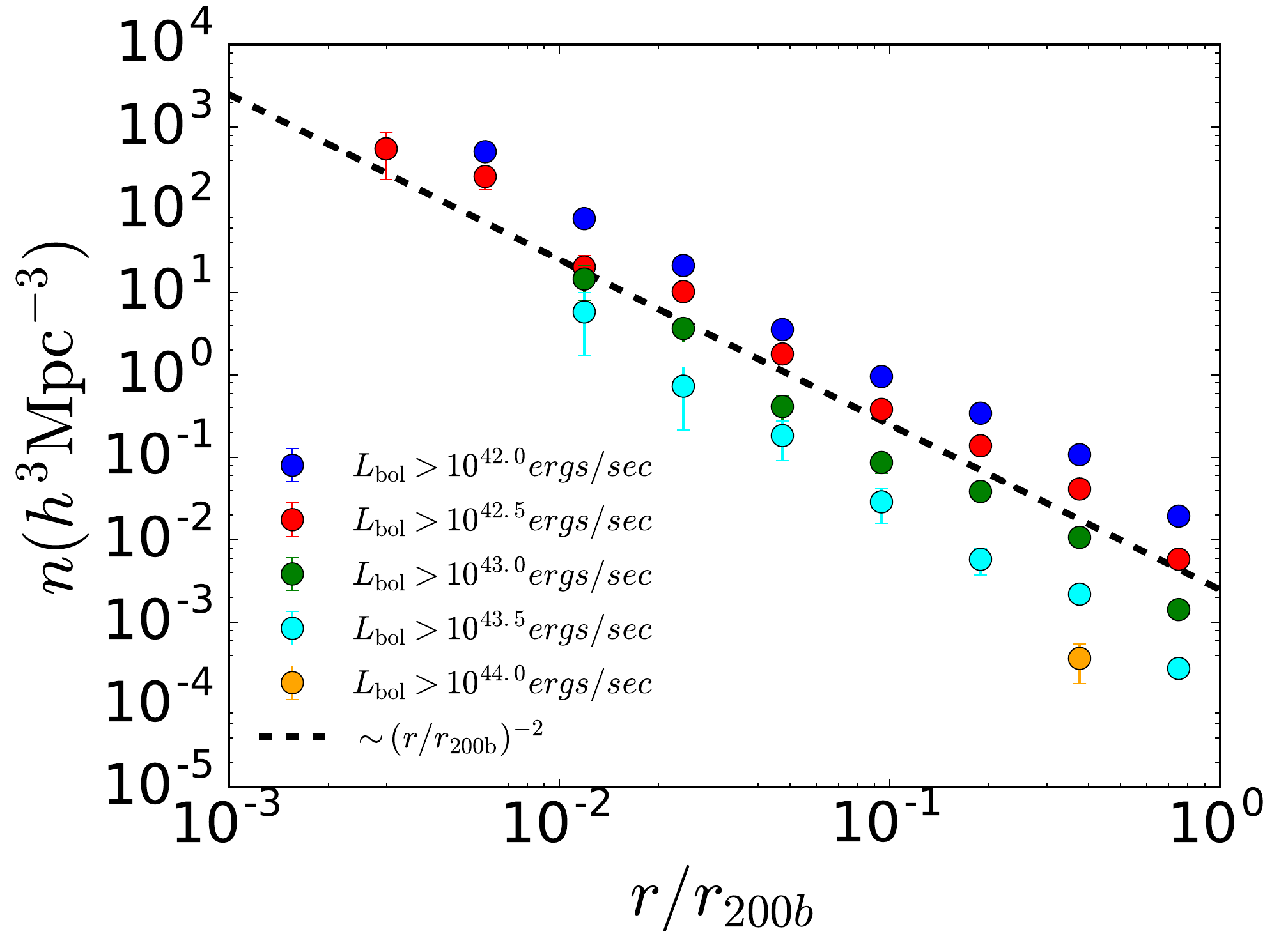} 
\caption{Radial satellite distributions at $z=1$ in haloes with $M_h>10^{11} M_{\odot}/h$. Different colors correspond to different bolometric luminosity thresholds. The dashed line is a power law with exponent -2 for reference. The radial distributions are consistent with a power law profile with exponent -2 at all redshifts over $0.6<z<1.5$.}
\label{radial_satellite_distributions_fig}
\end{figure}
For a given satellite occupation, the radial satellite distributions determine how the small-scale clustering power is distributed across different scales. 
Figure \ref{radial_satellite_distributions_fig} shows the radial profiles of satellite AGNs around the central AGN averaged over all haloes of mass $M_h>10^{11}~M_{\odot}/h$ at $z=1$. We find that across the entire range of luminosities (shown as different colors), profiles trace a power-law with exponent $\sim -2$. This is consistent with previous works on AGN HODs using smaller volume ($33.75~\mathrm{cMpc}/h$) hydrodynamic simulations \citep{2012MNRAS.419.2657C}. Therefore, for the modeling of the one-halo term, we use a satellite profile with exponent `-2'.

\begin{figure*}
\includegraphics[width=\textwidth]{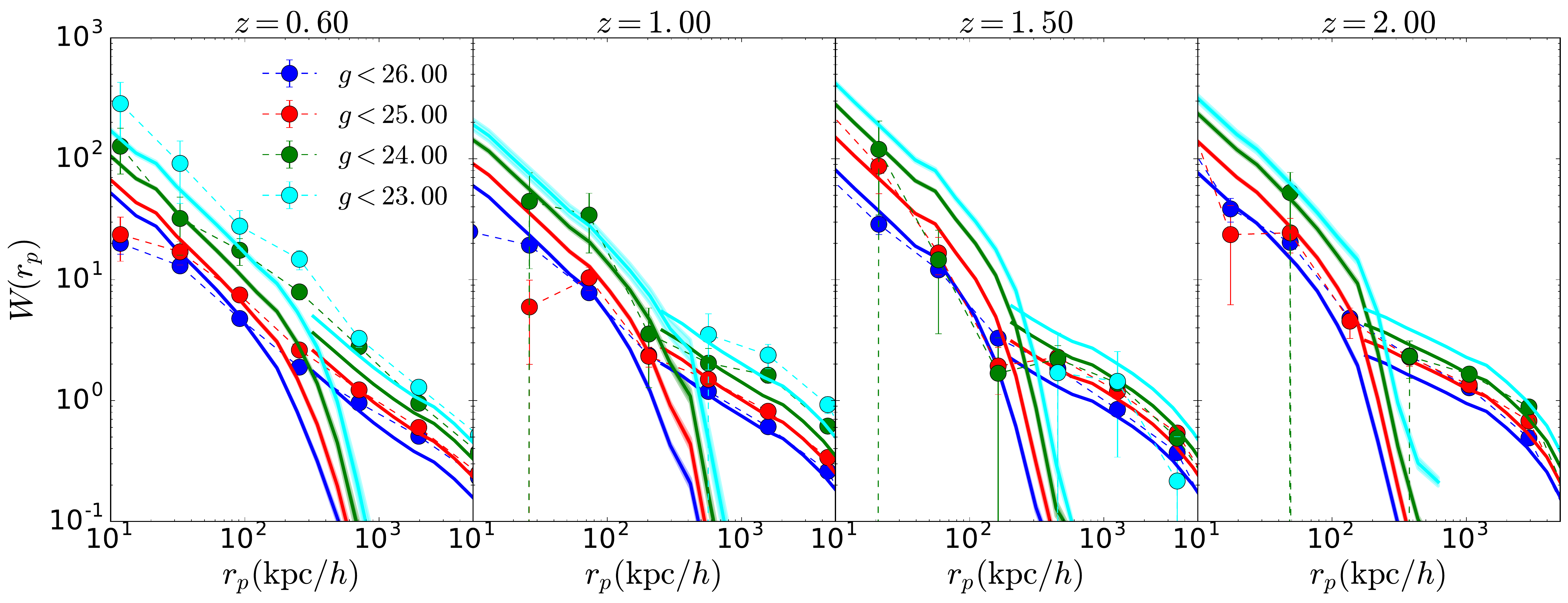}
\caption{Solid lines show the predicted one-halo and two-halo contributions of the projected clustering profiles of our CLF model summarized in Tables A1 and A2. The filled circles are the predictions from MBII.}
\label{projected_clustering_profile}
\end{figure*}

In order to validate our CLF model and our underlying assumptions, we compare the  projected clustering predictions against the direct predictions of our simulations for AGNs from $g<26$ up to $g<23$ (the regime that is well probed by our simulations). Figure \ref{projected_clustering_profile} shows the projected clustering profiles. The filled circles show the MBII predictions~(using Section \ref{determine_wp}: Method 1) and solid lines show the CLF model predictions~(using Section \ref{determine_wp}: Method 2). We can see that the simulations and CLF model predictions are within reasonable agreement for a wide range of redshifts and magnitude thresholds.

\bibliography{references}

\begin{thebibliography}{}
\makeatletter
\relax
\def\mn@urlcharsother{\let\do\@makeother \do\$\do\&\do\#\do\^\do\_\do\%\do\~}
\def\mn@doi{\begingroup\mn@urlcharsother \@ifnextchar [ {\mn@doi@}
  {\mn@doi@[]}}
\def\mn@doi@[#1]#2{\def\@tempa{#1}\ifx\@tempa\@empty \href
  {http://dx.doi.org/#2} {doi:#2}\else \href {http://dx.doi.org/#2} {#1}\fi
  \endgroup}
\def\mn@eprint#1#2{\mn@eprint@#1:#2::\@nil}
\def\mn@eprint@arXiv#1{\href {http://arxiv.org/abs/#1} {{\tt arXiv:#1}}}
\def\mn@eprint@dblp#1{\href {http://dblp.uni-trier.de/rec/bibtex/#1.xml}
  {dblp:#1}}
\def\mn@eprint@#1:#2:#3:#4\@nil{\def\@tempa {#1}\def\@tempb {#2}\def\@tempc
  {#3}\ifx \@tempc \@empty \let \@tempc \@tempb \let \@tempb \@tempa \fi \ifx
  \@tempb \@empty \def\@tempb {arXiv}\fi \@ifundefined
  {mn@eprint@\@tempb}{\@tempb:\@tempc}{\expandafter \expandafter \csname
  mn@eprint@\@tempb\endcsname \expandafter{\@tempc}}}

\bibitem[\protect\citeauthoryear{{Ballantyne}}{{Ballantyne}}{2017}]{2017MNRAS.464..613B}
{Ballantyne} D.~R.,  2017, \mn@doi [\mnras] {10.1093/mnras/stw2364}, \href
  {http://adsabs.harvard.edu/abs/2017MNRAS.464..613B} {464, 613}

\bibitem[\protect\citeauthoryear{{Chatterjee}, {Degraf}, {Richardson}, {Zheng},
  {Nagai}  \& {Di Matteo}}{{Chatterjee} et~al.}{2012}]{2012MNRAS.419.2657C}
{Chatterjee} S.,  {Degraf} C.,  {Richardson} J.,  {Zheng} Z.,  {Nagai} D.,
  {Di Matteo} T.,  2012, \mn@doi [\mnras] {10.1111/j.1365-2966.2011.19917.x},
  \href {http://adsabs.harvard.edu/abs/2012MNRAS.419.2657C} {419, 2657}

\bibitem[\protect\citeauthoryear{{Chatterjee}, {Nguyen}, {Myers}  \&
  {Zheng}}{{Chatterjee} et~al.}{2013}]{2013ApJ...779..147C}
{Chatterjee} S.,  {Nguyen} M.~L.,  {Myers} A.~D.,   {Zheng} Z.,  2013, \mn@doi
  [\apj] {10.1088/0004-637X/779/2/147}, \href
  {http://adsabs.harvard.edu/abs/2013ApJ...779..147C} {779, 147}

\bibitem[\protect\citeauthoryear{{Cooray}}{{Cooray}}{2006}]{2006MNRAS.365..842C}
{Cooray} A.,  2006, \mn@doi [\mnras] {10.1111/j.1365-2966.2005.09747.x}, \href
  {http://adsabs.harvard.edu/abs/2006MNRAS.365..842C} {365, 842}

\bibitem[\protect\citeauthoryear{{Croom} et~al.,}{{Croom}
  et~al.}{2005}]{2005MNRAS.356..415C}
{Croom} S.~M.,  et~al., 2005, \mn@doi [\mnras]
  {10.1111/j.1365-2966.2004.08379.x}, \href
  {http://adsabs.harvard.edu/abs/2005MNRAS.356..415C} {356, 415}

\bibitem[\protect\citeauthoryear{{Croom} et~al.,}{{Croom}
  et~al.}{2009}]{2009MNRAS.399.1755C}
{Croom} S.~M.,  et~al., 2009, \mn@doi [\mnras]
  {10.1111/j.1365-2966.2009.15398.x}, \href
  {http://adsabs.harvard.edu/abs/2009MNRAS.399.1755C} {399, 1755}

\bibitem[\protect\citeauthoryear{{Davis}, {Efstathiou}, {Frenk}  \&
  {White}}{{Davis} et~al.}{1985}]{1985ApJ...292..371D}
{Davis} M.,  {Efstathiou} G.,  {Frenk} C.~S.,   {White} S.~D.~M.,  1985,
  \mn@doi [\apj] {10.1086/163168}, \href
  {http://adsabs.harvard.edu/abs/1985ApJ...292..371D} {292, 371}

\bibitem[\protect\citeauthoryear{{Dawson} et~al.,}{{Dawson}
  et~al.}{2016}]{2016AJ....151...44D}
{Dawson} K.~S.,  et~al., 2016, \mn@doi [\aj] {10.3847/0004-6256/151/2/44},
  \href {http://adsabs.harvard.edu/abs/2016AJ....151...44D} {151, 44}

\bibitem[\protect\citeauthoryear{{DeGraf} \& {Sijacki}}{{DeGraf} \&
  {Sijacki}}{2017}]{2017MNRAS.466.3331D}
{DeGraf} C.,  {Sijacki} D.,  2017, \mn@doi [\mnras] {10.1093/mnras/stw3267},
  \href {http://adsabs.harvard.edu/abs/2017MNRAS.466.3331D} {466, 3331}

\bibitem[\protect\citeauthoryear{{Di Matteo}, {Springel}  \& {Hernquist}}{{Di
  Matteo} et~al.}{2005}]{2005Natur.433..604D}
{Di Matteo} T.,  {Springel} V.,   {Hernquist} L.,  2005, \mn@doi [\nat]
  {10.1038/nature03335}, \href
  {http://adsabs.harvard.edu/abs/2005Natur.433..604D} {433, 604}

\bibitem[\protect\citeauthoryear{{Di Matteo}, {Khandai}, {DeGraf}, {Feng},
  {Croft}, {Lopez}  \& {Springel}}{{Di Matteo}
  et~al.}{2012}]{2012ApJ...745L..29D}
{Di Matteo} T.,  {Khandai} N.,  {DeGraf} C.,  {Feng} Y.,  {Croft} R.~A.~C.,
  {Lopez} J.,   {Springel} V.,  2012, \mn@doi [\apjl]
  {10.1088/2041-8205/745/2/L29}, \href
  {http://adsabs.harvard.edu/abs/2012ApJ...745L..29D} {745, L29}

\bibitem[\protect\citeauthoryear{{Eftekharzadeh} et~al.,}{{Eftekharzadeh}
  et~al.}{2015}]{2015MNRAS.453.2779E}
{Eftekharzadeh} S.,  et~al., 2015, \mn@doi [\mnras] {10.1093/mnras/stv1763},
  \href {http://adsabs.harvard.edu/abs/2015MNRAS.453.2779E} {453, 2779}

\bibitem[\protect\citeauthoryear{{Eftekharzadeh}, {Myers}, {Hennawi},
  {Djorgovski}, {Richards}, {Mahabal}  \& {Graham}}{{Eftekharzadeh}
  et~al.}{2017}]{2017MNRAS.468...77E}
{Eftekharzadeh} S.,  {Myers} A.~D.,  {Hennawi} J.~F.,  {Djorgovski} S.~G.,
  {Richards} G.~T.,  {Mahabal} A.~A.,   {Graham} M.~J.,  2017, \mn@doi [\mnras]
  {10.1093/mnras/stx412}, \href
  {http://adsabs.harvard.edu/abs/2017MNRAS.468...77E} {468, 77}

\bibitem[\protect\citeauthoryear{{Eftekharzadeh}, {Myers}  \&
  {Kourkchi}}{{Eftekharzadeh} et~al.}{2018}]{2018arXiv181205760E}
{Eftekharzadeh} S.,  {Myers} A.~D.,   {Kourkchi} E.,  2018, arXiv e-prints,
  \href {http://adsabs.harvard.edu/abs/2018arXiv181205760E} {}

\bibitem[\protect\citeauthoryear{{Fakhouri}, {Ma}  \&
  {Boylan-Kolchin}}{{Fakhouri} et~al.}{2010}]{2010MNRAS.406.2267F}
{Fakhouri} O.,  {Ma} C.-P.,   {Boylan-Kolchin} M.,  2010, \mn@doi [\mnras]
  {10.1111/j.1365-2966.2010.16859.x}, \href
  {http://adsabs.harvard.edu/abs/2010MNRAS.406.2267F} {406, 2267}

\bibitem[\protect\citeauthoryear{{Guo} et~al.,}{{Guo}
  et~al.}{2016}]{2016MNRAS.459.3040G}
{Guo} H.,  et~al., 2016, \mn@doi [\mnras] {10.1093/mnras/stw845}, \href
  {http://adsabs.harvard.edu/abs/2016MNRAS.459.3040G} {459, 3040}

\bibitem[\protect\citeauthoryear{{Harikane} et~al.,}{{Harikane}
  et~al.}{2018}]{2018PASJ...70S..11H}
{Harikane} Y.,  et~al., 2018, \mn@doi [\pasj] {10.1093/pasj/psx097}, \href
  {http://adsabs.harvard.edu/abs/2018PASJ...70S..11H} {70, S11}

\bibitem[\protect\citeauthoryear{{Hennawi} et~al.,}{{Hennawi}
  et~al.}{2006}]{2006AJ....131....1H}
{Hennawi} J.~F.,  et~al., 2006, \mn@doi [\aj] {10.1086/498235}, \href
  {http://adsabs.harvard.edu/abs/2006AJ....131....1H} {131, 1}

\bibitem[\protect\citeauthoryear{{Kaviraj} et~al.,}{{Kaviraj}
  et~al.}{2017}]{2017MNRAS.467.4739K}
{Kaviraj} S.,  et~al., 2017, \mn@doi [\mnras] {10.1093/mnras/stx126}, \href
  {http://adsabs.harvard.edu/abs/2017MNRAS.467.4739K} {467, 4739}

\bibitem[\protect\citeauthoryear{{Kayo} \& {Oguri}}{{Kayo} \&
  {Oguri}}{2012}]{2012MNRAS.424.1363K}
{Kayo} I.,  {Oguri} M.,  2012, \mn@doi [\mnras]
  {10.1111/j.1365-2966.2012.21321.x}, \href
  {http://adsabs.harvard.edu/abs/2012MNRAS.424.1363K} {424, 1363}

\bibitem[\protect\citeauthoryear{{Khandai}, {Di Matteo}, {Croft}, {Wilkins},
  {Feng}, {Tucker}, {DeGraf}  \& {Liu}}{{Khandai}
  et~al.}{2015}]{2015MNRAS.450.1349K}
{Khandai} N.,  {Di Matteo} T.,  {Croft} R.,  {Wilkins} S.,  {Feng} Y.,
  {Tucker} E.,  {DeGraf} C.,   {Liu} M.-S.,  2015, \mn@doi [\mnras]
  {10.1093/mnras/stv627}, \href
  {http://adsabs.harvard.edu/abs/2015MNRAS.450.1349K} {450, 1349}

\bibitem[\protect\citeauthoryear{{Kochanek}, {Falco}  \&
  {Mu{\~n}oz}}{{Kochanek} et~al.}{1999}]{1999ApJ...510..590K}
{Kochanek} C.~S.,  {Falco} E.~E.,   {Mu{\~n}oz} J.~A.,  1999, \mn@doi [\apj]
  {10.1086/306594}, \href {http://adsabs.harvard.edu/abs/1999ApJ...510..590K}
  {510, 590}

\bibitem[\protect\citeauthoryear{{Komatsu} et~al.,}{{Komatsu}
  et~al.}{2011}]{2011ApJS..192...18K}
{Komatsu} E.,  et~al., 2011, \mn@doi [\apjs] {10.1088/0067-0049/192/2/18},
  \href {http://adsabs.harvard.edu/abs/2011ApJS..192...18K} {192, 18}

\bibitem[\protect\citeauthoryear{{Leauthaud} et~al.,}{{Leauthaud}
  et~al.}{2015}]{2015MNRAS.446.1874L}
{Leauthaud} A.,  et~al., 2015, \mn@doi [\mnras] {10.1093/mnras/stu2210}, \href
  {http://adsabs.harvard.edu/abs/2015MNRAS.446.1874L} {446, 1874}

\bibitem[\protect\citeauthoryear{{McGreer} et~al.,}{{McGreer}
  et~al.}{2013}]{2013ApJ...768..105M}
{McGreer} I.~D.,  et~al., 2013, \mn@doi [\apj] {10.1088/0004-637X/768/2/105},
  \href {http://adsabs.harvard.edu/abs/2013ApJ...768..105M} {768, 105}

\bibitem[\protect\citeauthoryear{{McGreer}, {Eftekharzadeh}, {Myers}  \&
  {Fan}}{{McGreer} et~al.}{2016}]{2016AJ....151...61M}
{McGreer} I.~D.,  {Eftekharzadeh} S.,  {Myers} A.~D.,   {Fan} X.,  2016,
  \mn@doi [\aj] {10.3847/0004-6256/151/3/61}, \href
  {http://adsabs.harvard.edu/abs/2016AJ....151...61M} {151, 61}

\bibitem[\protect\citeauthoryear{{Mitra}, {Chatterjee}, {DiPompeo}, {Myers}  \&
  {Zheng}}{{Mitra} et~al.}{2018}]{2018MNRAS.477...45M}
{Mitra} K.,  {Chatterjee} S.,  {DiPompeo} M.~A.,  {Myers} A.~D.,   {Zheng} Z.,
  2018, \mn@doi [\mnras] {10.1093/mnras/sty556}, \href
  {http://adsabs.harvard.edu/abs/2018MNRAS.477...45M} {477, 45}

\bibitem[\protect\citeauthoryear{{Mortlock}, {Webster}  \&
  {Francis}}{{Mortlock} et~al.}{1999}]{1999MNRAS.309..836M}
{Mortlock} D.~J.,  {Webster} R.~L.,   {Francis} P.~J.,  1999, \mn@doi [\mnras]
  {10.1046/j.1365-8711.1999.02872.x}, \href
  {http://adsabs.harvard.edu/abs/1999MNRAS.309..836M} {309, 836}

\bibitem[\protect\citeauthoryear{{Myers} et~al.,}{{Myers}
  et~al.}{2006}]{2006ApJ...638..622M}
{Myers} A.~D.,  et~al., 2006, \mn@doi [\apj] {10.1086/499093}, \href
  {http://adsabs.harvard.edu/abs/2006ApJ...638..622M} {638, 622}

\bibitem[\protect\citeauthoryear{{Myers}, {Brunner}, {Richards}, {Nichol},
  {Schneider}  \& {Bahcall}}{{Myers} et~al.}{2007}]{2007ApJ...658...99M}
{Myers} A.~D.,  {Brunner} R.~J.,  {Richards} G.~T.,  {Nichol} R.~C.,
  {Schneider} D.~P.,   {Bahcall} N.~A.,  2007, \mn@doi [\apj] {10.1086/511520},
  \href {http://adsabs.harvard.edu/abs/2007ApJ...658...99M} {658, 99}

\bibitem[\protect\citeauthoryear{{Myers}, {Richards}, {Brunner}, {Schneider},
  {Strand}, {Hall}, {Blomquist}  \& {York}}{{Myers}
  et~al.}{2008}]{2008ApJ...678..635M}
{Myers} A.~D.,  {Richards} G.~T.,  {Brunner} R.~J.,  {Schneider} D.~P.,
  {Strand} N.~E.,  {Hall} P.~B.,  {Blomquist} J.~A.,   {York} D.~G.,  2008,
  \mn@doi [\apj] {10.1086/533491}, \href
  {http://adsabs.harvard.edu/abs/2008ApJ...678..635M} {678, 635}

\bibitem[\protect\citeauthoryear{{Myers} et~al.,}{{Myers}
  et~al.}{2015}]{2015ApJS..221...27M}
{Myers} A.~D.,  et~al., 2015, \mn@doi [\apjs] {10.1088/0067-0049/221/2/27},
  \href {http://adsabs.harvard.edu/abs/2015ApJS..221...27M} {221, 27}

\bibitem[\protect\citeauthoryear{{Nelson} et~al.,}{{Nelson}
  et~al.}{2015}]{2015A&C....13...12N}
{Nelson} D.,  et~al., 2015, \mn@doi [Astronomy and Computing]
  {10.1016/j.ascom.2015.09.003}, \href
  {http://adsabs.harvard.edu/abs/2015A%26C....13...12N} {13, 12}

\bibitem[\protect\citeauthoryear{{Palanque-Delabrouille}
  et~al.,}{{Palanque-Delabrouille} et~al.}{2016}]{2016A&A...587A..41P}
{Palanque-Delabrouille} N.,  et~al., 2016, \mn@doi [\aap]
  {10.1051/0004-6361/201527392}, \href
  {https://ui.adsabs.harvard.edu/\#abs/2016A&A...587A..41P} {587, A41}

\bibitem[\protect\citeauthoryear{{Porciani} \& {Norberg}}{{Porciani} \&
  {Norberg}}{2006}]{2006MNRAS.371.1824P}
{Porciani} C.,  {Norberg} P.,  2006, \mn@doi [\mnras]
  {10.1111/j.1365-2966.2006.10813.x}, \href
  {http://adsabs.harvard.edu/abs/2006MNRAS.371.1824P} {371, 1824}

\bibitem[\protect\citeauthoryear{{Richards} et~al.,}{{Richards}
  et~al.}{2006}]{2006AJ....131.2766R}
{Richards} G.~T.,  et~al., 2006, \mn@doi [\aj] {10.1086/503559}, \href
  {http://adsabs.harvard.edu/abs/2006AJ....131.2766R} {131, 2766}

\bibitem[\protect\citeauthoryear{{Richardson}, {Zheng}, {Chatterjee}, {Nagai}
  \& {Shen}}{{Richardson} et~al.}{2012}]{2012ApJ...755...30R}
{Richardson} J.,  {Zheng} Z.,  {Chatterjee} S.,  {Nagai} D.,   {Shen} Y.,
  2012, \mn@doi [\apj] {10.1088/0004-637X/755/1/30}, \href
  {http://adsabs.harvard.edu/abs/2012ApJ...755...30R} {755, 30}

\bibitem[\protect\citeauthoryear{{Schaye} et~al.,}{{Schaye}
  et~al.}{2015}]{2015MNRAS.446..521S}
{Schaye} J.,  et~al., 2015, \mn@doi [\mnras] {10.1093/mnras/stu2058}, \href
  {http://adsabs.harvard.edu/abs/2015MNRAS.446..521S} {446, 521}

\bibitem[\protect\citeauthoryear{{Schneider} et~al.,}{{Schneider}
  et~al.}{2000}]{2000AJ....120.2183S}
{Schneider} D.~P.,  et~al., 2000, \mn@doi [\aj] {10.1086/316834}, \href
  {http://adsabs.harvard.edu/abs/2000AJ....120.2183S} {120, 2183}

\bibitem[\protect\citeauthoryear{{Shen}}{{Shen}}{2013}]{2013BASI...41...61S}
{Shen} Y.,  2013, Bulletin of the Astronomical Society of India, \href
  {http://adsabs.harvard.edu/abs/2013BASI...41...61S} {41, 61}

\bibitem[\protect\citeauthoryear{{Shen} et~al.,}{{Shen}
  et~al.}{2009}]{2009ApJ...697.1656S}
{Shen} Y.,  et~al., 2009, \mn@doi [\apj] {10.1088/0004-637X/697/2/1656}, \href
  {http://adsabs.harvard.edu/abs/2009ApJ...697.1656S} {697, 1656}

\bibitem[\protect\citeauthoryear{{Springel}}{{Springel}}{2005}]{2005MNRAS.364.1105S}
{Springel} V.,  2005, \mn@doi [\mnras] {10.1111/j.1365-2966.2005.09655.x},
  \href {http://adsabs.harvard.edu/abs/2005MNRAS.364.1105S} {364, 1105}

\bibitem[\protect\citeauthoryear{{Springel} \& {Hernquist}}{{Springel} \&
  {Hernquist}}{2003}]{2003MNRAS.339..289S}
{Springel} V.,  {Hernquist} L.,  2003, \mn@doi [\mnras]
  {10.1046/j.1365-8711.2003.06206.x}, \href
  {http://adsabs.harvard.edu/abs/2003MNRAS.339..289S} {339, 289}

\bibitem[\protect\citeauthoryear{{Springel}, {Di Matteo}  \&
  {Hernquist}}{{Springel} et~al.}{2005}]{2005MNRAS.361..776S}
{Springel} V.,  {Di Matteo} T.,   {Hernquist} L.,  2005, \mn@doi [\mnras]
  {10.1111/j.1365-2966.2005.09238.x}, \href
  {http://adsabs.harvard.edu/abs/2005MNRAS.361..776S} {361, 776}

\bibitem[\protect\citeauthoryear{{Tinker}, {Kravtsov}, {Klypin}, {Abazajian},
  {Warren}, {Yepes}, {Gottl{\"o}ber}  \& {Holz}}{{Tinker}
  et~al.}{2008}]{2008ApJ...688..709T}
{Tinker} J.,  {Kravtsov} A.~V.,  {Klypin} A.,  {Abazajian} K.,  {Warren} M.,
  {Yepes} G.,  {Gottl{\"o}ber} S.,   {Holz} D.~E.,  2008, \mn@doi [\apj]
  {10.1086/591439}, \href {http://adsabs.harvard.edu/abs/2008ApJ...688..709T}
  {688, 709}

\bibitem[\protect\citeauthoryear{{Tinker}, {Robertson}, {Kravtsov}, {Klypin},
  {Warren}, {Yepes}  \& {Gottl{\"o}ber}}{{Tinker}
  et~al.}{2010}]{2010ApJ...724..878T}
{Tinker} J.~L.,  {Robertson} B.~E.,  {Kravtsov} A.~V.,  {Klypin} A.,  {Warren}
  M.~S.,  {Yepes} G.,   {Gottl{\"o}ber} S.,  2010, \mn@doi [\apj]
  {10.1088/0004-637X/724/2/878}, \href
  {http://adsabs.harvard.edu/abs/2010ApJ...724..878T} {724, 878}

\bibitem[\protect\citeauthoryear{{Trevisan} \& {Mamon}}{{Trevisan} \&
  {Mamon}}{2017}]{2017MNRAS.471.2022T}
{Trevisan} M.,  {Mamon} G.~A.,  2017, \mn@doi [\mnras] {10.1093/mnras/stx1656},
  \href {http://adsabs.harvard.edu/abs/2017MNRAS.471.2022T} {471, 2022}

\bibitem[\protect\citeauthoryear{{White} et~al.,}{{White}
  et~al.}{2012}]{2012MNRAS.424..933W}
{White} M.,  et~al., 2012, \mn@doi [\mnras] {10.1111/j.1365-2966.2012.21251.x},
  \href {http://adsabs.harvard.edu/abs/2012MNRAS.424..933W} {424, 933}

\bibitem[\protect\citeauthoryear{{Zheng}, {Coil}  \& {Zehavi}}{{Zheng}
  et~al.}{2007}]{2007ApJ...667..760Z}
{Zheng} Z.,  {Coil} A.~L.,   {Zehavi} I.,  2007, \mn@doi [\apj]
  {10.1086/521074}, \href {http://adsabs.harvard.edu/abs/2007ApJ...667..760Z}
  {667, 760}

\makeatother
\end{thebibliography}
\end{document}